\documentclass{jstatphys}

\usepackage[reqno,sumlimits]{amsmath} \usepackage{amsfonts,amssymb,amsbsy}
\usepackage{isolatin1,submitextras}

\newif\ifarXiv      

\makeatletter\@ifundefined{comp}{\arXivtrue}{\arXivfalse}\makeatother

\ifarXiv
\else
\fi

\begin{document}

\ifarXiv
  \title[Spectral Localization by Gaussian Random Potentials]%
          {Spectral Localization by\\ Gaussian Random
           Potentials in\\[1ex] Multi-Dimensional Continuous Space} 
\else
  \title[Spectral Localization by Gaussian Random Potentials]%
        {Spectral Localization by Gaussian Random\\[0.6ex]
         Potentials in Multi-Dimensional Continuous Space} 
\fi

\headnote{To appear in slightly different form in Journal of
  Statistical Physics}

\author{Werner Fischer,%
             \footnote{Institut f\"ur Theoretische Physik, 
                       Universit\"at  
                       Erlangen-N\"urnberg, Staudtstra{\ss}e 7, D--91058 
                       Erlangen, Germany.
                       E-mail: leschke@theorie1.physik.uni-erlangen.de 
                       .}$^{,\,\!}$%
             \footnote{New address: Infineon Technologies,
                       P.\ O.\ Box 800949, D--81609 M\"unchen, Germany.}
        Hajo Leschke,\footnotemark[1] 
        and Peter M\"uller%
             \footnote{Institut f\"ur Theoretische Physik, 
                       Georg-August-Universit\"at, \mbox{D--37073} 
                       G\"ot\-ting\-en, Germany. 
                       E-mail: Peter.Mueller@Physik.Uni-Goettingen.DE .}%
       }
\runningauthor{Fischer, Leschke, M\"uller}

\date{12 October 2000, ~final version of math-ph/9912025}

\begin{abstract}
A detailed mathematical proof is given that the energy spectrum of a
non-relativistic quantum particle  
in multi-dimensional Euclidean space under the influence of  
suitable random potentials has almost surely a pure-point component.
The result applies in particular to a certain class of zero-mean 
Gaussian random potentials, which are homogeneous with respect to
Euclidean translations. More precisely, for these Gaussian random
potentials the spectrum is almost surely only pure point at
sufficiently negative energies or, at negative energies, for
sufficiently weak disorder. 
The proof is based on a fixed-energy multi-scale
analysis which allows for different random potentials on different
length scales.
\end{abstract}


\medskip\noindent
\emph{Dedicated to the memory of Uwe Brandt (15 April 1944 ~--~
  1 November 1997).}
\smallskip


%
\tableofcontents
\section{Introduction} 

Even more than 40 years after Anderson's pioneering paper \cite{And58},
single-particle Schr\"odinger operators with random potentials
continue to
play a prominent r\^ole for understanding the suppression of charge transport
in disordered solids by electronic localization. In this context
both spectral and dynamical criteria for localization are commonly
studied. Spectral localization means that there is only (dense) pure-point
spectrum in certain energy regimes. In addition, the corresponding
eigenfunctions are often required to decay exponentially at
infinity. As to criteria for dynamical localization, we mention
a sufficiently slow long-time growth of the spreading of
quantum states or the vanishing of the direct-current
conductivity of the corresponding ideal Fermi gas at zero temperature;
see \cite{FrSp83,MaHo84,GeDBi98,AiGr98,BaFi99,DaSt99} for works along
these lines. Yet, the
interrelations between the different criteria 
are  not understood in sufficient generality and are presently under
active debate \cite{Sim90, DRJi96, Las96, BaCo97c, KiLa99}.

The goal of this paper is to contribute to the understanding of
spectral localization for random Schr\"odinger operators in
multi-dimensional Euclidean space $\rz^{d}$. 
Using physical units in which the
mass of the particle, its electric charge and Planck's constant divided by
$2\pi$ are all equal to one, these
operators are informally given by differential expressions of the form
\begin{equation} \label{formalH}
  H(V) = \frac{1}{2}\,(\i\nabla +a)^{2} +V\,.
\end{equation}
Here, $\i=\sqrt{-1}$ is the imaginary unit and $\nabla$ is the nabla
operator. The
non-random vector potential $a$ allows for the presence of a
magnetic field $\nabla\times a$, and the scalar potential $V$ is an ergodic
random field. We are primarily interested in a constant magnetic
field and a zero-mean Gaussian random potential $V$. Under certain
assumptions on the covariance function of $V$ we will show that the
spectrum of $H(V)$ is almost surely only pure point at
sufficiently negative energies or, at negative energies, for
sufficiently weak disorder, 
see Theorem~\ref{maintheo} below. In the physics 
literature such a result has been inferred from non-rigorous arguments
several decades ago and is nowadays usually taken for granted. These arguments
rely on the idea that at sufficiently low energies even weak
disorder should be able to suppress quantum-mechanical tunnelling. Our point
here is to present a complete and detailed mathematical proof.

The first mathematical proof of spectral localization for arbitrary space
dimensions $d\ge 1$ dates back to \cite{FrSp83}, who considered
Anderson's original model on the simple cubic lattice
$\mathbb{Z}^{d}$. Their method of
proof is a so-called multi-scale analysis, where elements from
Kolmogorov-Arnold-Moser
theory are invoked for coping  with small denominators in order to
bound resolvents of finite-volume random Schr\"odinger operators
with high probabilities. Consequences of this method were elaborated on in 
\cite{MaSc85,DeLe85,FrMa85,SiWo86}. Using scaling ideas borrowed from
percolation 
theory, a substantial simplification and streamlining of the
multi-scale analysis is due to \cite{Spe88,DrKl89}. Another
breakthrough in proving spectral localization for lattice models was
\cite{AiMo93} where a completely different, technically much
less involved proof was presented; see also
\cite{Aiz94,Gra94,Hun97,AiGr98,DoMa99,AiSc99}
for subsequent works 
along these lines. In contrast to the multi-scale analysis the method of 
\cite{AiMo93} does not seem to be adaptable to general continuum
models as \eqref{formalH}. 

Notably, the first 
proof \cite{MaHo84} of spectral localization for a random
Schr\"odinger operator in multi-dimensional continuous space $\rz^{d}$
appeared shortly 
after \cite{FrSp83}. Yet, it took ten more years until these
investigations were continued in \cite{CoHi94}, a paper which opened
the field for a deeper understanding of Schr\"odinger operators with
random alloy-type potentials in multi-dimensional continuous space.
Spectral localization is now known to occur near the band edges of the
spectrum of randomly perturbed periodic Schr\"odinger operators
\cite{Klo95a,BaCo97b,KiSt98a,KiSt98b,Ves98,Sto00} and near the band edges of
disorder-broadened Landau levels arising from operators of the form
\eqref{formalH} \cite{CoHi96,BaCo97a,Wan97,DoMa99}. Additional results
are available 
when restricting the latter onto the eigenspace of a single Landau level
\cite{DoMa95,DoMa96,PuSc97,Scr99}. In \cite{Klo95b} spectral
localization is established near the bottom of the spectrum, but without
an otherwise frequently used positivity assumption on the single-site
potential, see also \cite{BuSt98} for similar results for one dimension. 
Models with random point interactions allow for more specific methods
in order to prove spectral localization \cite{BoGr97,DoMa99,Scr99}.
Good overviews of the mathematical theory related to Anderson
localization can be found in the survey articles
\cite{Spe86,MaSc87,Kir89} and the monographs \cite{CaLa90,PaFi92,Sto00}. 

To our knowledge, proofs of spectral
localization in multi-dimensional continuous space $\rz^{d}$ have so
far been completed only for alloy-type random potentials and related
ones. Basically, these potentials still have an underlying lattice
structure and, in some works \cite{KiSt98a,Zen99,Sto00}, were allowed
to exhibit correlations via a long-range tail of the single-site potential.
In contrast, the methods developed in this paper are tailored
for a wide class of truly continuum random potentials in
multi-dimensional continuous space, which -- as is the case for
Gaussian random potentials -- may have
unbounded fluctuations and genuine long-range 
correlations. Besides mathematical challenges, the motivation for
coping with these additional 
difficulties stems from the fact that Gaussian random potentials are
appealing for at least three reasons. 
First, there is a belief in the ``normality of the normal
distribution'' in nature. Second, the $n$-point cumulant functions
of Gaussian random potentials vanish for all $n\ge 3$, a fact which
leads to computational simplifications. Third, the degree of
randomness can be varied by choosing different covariance functions,
that is, $2$-point cumulant functions. For all three reasons Gaussian
random potentials find widespread applications in -- if not dominate --
the corresponding physics literature, see for example the books
\cite{BoEn84,ShEf84,LiGr88,Efe97} and references therein. 

Due to the long-range correlations of Gaussian random potentials we
were not able to perform a ``variable-energy'' multi-scale analysis in
the spirit of \cite{Spe88,DrKl89} in order to prove localization. Instead we
build on the ``fixed-energy'' multi-scale analysis of \cite{DrKl91},
which guarantees that all events, whose joint probability has to be
estimated, are far enough apart. As a consequence, we obtain only
algebraic (and not exponential) decay estimates. The necessary
techniques for coping with 
continuum Schr\"odinger operators are patterned after \cite{CoHi94}.
We also refer to the remarks at the beginning of
Section~\ref{multiscale} for a brief description of the multi-scale
analysis used in this paper. 

Part of the localization result, whose proof is presented here in
full detail, was
announced in \cite{FiLe96} and \cite{FiHu97a}. Its key ideas and the
way in which the assumptions on the Gaussian random potential enter
were briefly outlined in \cite{FiLe96}, as far as the absence of the
absolutely continuous spectrum is concerned. An important
ingredient of the proof, a so-called Wegner estimate for Schr\"odinger
operators with Gaussian random potentials, was formulated and proven
in \cite{FiHu97b}. It is the purpose of this paper to provide the 
remaining details for a complete proof of the absence of both the
absolutely continuous and the singular continuous spectrum. 
One reason why we have decided to give a rather explicit exposition is
that our proof of spectral localization requires stronger assumptions
on the Gaussian random potential than those being sufficient for
obtaining the Wegner estimate. Another reason is that Stollmann's
recent book \cite{Sto00} -- the only source we are aware of which
provides a detailed multi-scale analysis for continuum models with
long-range correlated random potentials -- does not cover the type of
random potentials considered here.

The plan of the paper is as follows. Section~\ref{schroeder} serves to
fix the basic notation and to give a precise
definition of the random Schr\"odinger operators we are interested in. In 
Theorem~\ref{maintheo} of Subsection~\ref{resultsection} we state our
localization result for Schr\"odinger operators with Gaussian random
potentials. In Subsection \ref{wegnerSec} we recall the Wegner
estimate from \cite{FiHu97b}.
The multi-scale analysis is presented in detail in Section~\ref{multiscale}. 
All results of Sections~\ref{multiscale}
and~\ref{ppSec} are formulated and proven for a rather general
class of truly continuum correlated random potentials, which includes
Gaussian ones. 
The main technical result of the multi-scale analysis, resolvent
estimates on multiple length scales, is formulated in 
Subsection~\ref{finvol} as Theorem~\ref{multThm}. The transition to
the infinite-volume resolvent is performed in Subsection~\ref{infvol}.
In Section~\ref{ppSec} we show how to build on the results from
Section~\ref{multiscale} in order to conclude that the spectrum is
only pure point in certain energy regimes. 
Section~\ref{applSec} is devoted to
the proof of the main Theorem~\ref{maintheo} for Gaussian
random potentials. This is done by verifying that the more general
theorems of Sections~\ref{multiscale} and~\ref{ppSec} can be applied.
In the Appendix we present a simple explicit Combes-Thomas type of estimate
needed in Subsection~\ref{inessec}.


\section{Basic Definitions and Main Result} \label{schroeder}

In Subsection~\ref{defsection} we fix our notation, give a precise
definition of random Schr\"odinger operators and compile some basic
facts. Our main result is stated in Subsection~\ref{resultsection}.
Finally, we recall a Wegner estimate in Subsection~\ref{wegnerSec}.

\subsection{Random Schr\"odinger Operators} \label{defsection}

As usual, let $\bbN := \{1,2,3,\ldots\}$ denote the set of natural
numbers and $\bbN_{\,0}:=\bbN\cup \{0\}$. Let $\rz$, respectively $\cz$,
denote the field of real, respectively complex, numbers and set 
$\rz^{+}:=\{r\in\rz : r>0\}$ and $\rz^{+}_{\,0}:= \rz^{+}\cup\{0\}$. An
open cube $\Lambda$ in $d$-dimensional Euclidean space $\rz^{d}$,
$d\in\nz$, is 
the $d$-fold Cartesian product $\Lambda := I\times\ldots\times I$ of
an open interval $I\subseteq\rz$, and $\partial\Lambda$ stands for the
boundary of $\Lambda$. The open
cube with edges of length $l>0$ and centre $x\in\rz^{d}$  is the set
$\Lambda_{l}(x):= \{y\in\rz^{d}: |x-y|_{\infty}< l/2\}$. Here
$|x|_{\infty} := \max_{j=1,\ldots,d}|x_{j}|$ denotes the maximum norm
of $x=(x_{1},\ldots,x_{d})\in\rz^{d}$. The Euclidean scalar
product $x\cdot y:= \sum_{j=1}^{d}x_{j}y_{j}$ of $x,y\in\rz^{d}$
induces the Euclidean norm $|x|:=(x\cdot x)^{1/2}$ of
$x\in\rz^{d}$. We also write $x^{2}:=x\cdot x$. The distance of 
two subsets $\Lambda,\Lambda'\subseteq\rz^{d}$ with respect to the
Euclidean (maximum) norm is defined as
$\dist_{(\infty)}(\Lambda,\Lambda') := \inf\{|x-y|_{(\infty)}:
x\in\Lambda, y\in\Lambda'\}$.

Given a Borel subset $\Lambda\subseteq\rz^{d}$ we denote its volume
with respect to the $d$-dimensional Lebesgue measure as $|\Lambda| :=
\int_{\Lambda}\!\d^{d}x$. The Banach space $\RL^{p}(\Lambda)$ consists
of the Borel measurable complex-valued functions $\varphi:
\rz^{d}\rightarrow\cz$ which are identified if their values differ only on a
set of Lebesgue measure zero and possess a finite norm 
\begin{equation}
  \norm[p;\Lambda]{\varphi} := \left\{
    \begin{array}{ll} \displaystyle
      \biggl(\int_{\Lambda}\!\d^{d}x\; |\varphi(x)|^{p}\biggr)^{1/p}\;
      & \text{if}~p<\infty\,,\\[2.5ex]
      \displaystyle\esssup\limits_{x\in\Lambda} |\varphi(x)|
      & \text{if}~p=\infty\,.
    \end{array}
  \right.
\end{equation}
We use the abbreviation
$\norm[p]{\varphi}:=\norm[p;\Lambda]{\varphi}$ if there is no ambiguity.
For $\varphi\in\RL^{2}(\Lambda)$ we also use the notation
$\norm{\varphi} 
:= \norm[2]{\varphi}$ and recall that $\RL^{2}(\Lambda)$ becomes a
separable Hilbert space when equipped with the scalar product 
\begin{equation} \label{scalarproduct}
  \langle\varphi,\psi\rangle := \int_{\Lambda}\d^{d}x\,
  \bigl(\varphi(x)\bigr)^{*}\,\psi(x) \,.
\end{equation}
Here the star denotes complex conjugation. We write $\varphi\in
\RL^{p}_{\text{loc}}(\rz^{d})$, if $\varphi\in\RL^{p}(\Lambda)$ for all 
$\Lambda\subset\rz^{d}$ with finite volume.
Finally, $\mathcal{C}^{n}(\rz^{d})$ stands for the vector space of
functions $\varphi: \rz^{d}\rightarrow\cz$ which are $n$ times continuously
differentiable and $\mathcal{C}^{\infty}_{0}(\rz^{d})$ for the vector
space of functions $\varphi: \rz^{d}\rightarrow\cz$ which are
arbitrarily often differentiable and have compact support. 

The norm of a bounded operator $A:\RL^{2}(\Lambda)\rightarrow
\RL^{2}(\Lambda)$  is defined as $\norm{A} := \sup\{\norm{A\varphi}:
\varphi\in\RL^{2}(\Lambda), \norm{\varphi}=1 \}$. The $d$-dimensional
gradient or nabla operator $\nabla:=(\partial/\partial x_{1},\ldots,
\partial/\partial x_{d})$ gives rise to the self-adjoint negative
Laplacian $-\Delta: \mathcal{D}(-\Delta)\ni\varphi\mapsto -
\nabla^{2}\varphi$ 
with domain $\mathcal{D}(-\Delta) := \big\{ \varphi\in\RL^{2}(\rz^{d}) :
\varphi, (\nabla\varphi)_{1},\ldots,(\nabla\varphi)_{d}$ are absolutely
continuous on $\rz^{d}$ and $\nabla^{2}\varphi\in\RL^{2}(\rz^{d})\big\}$.
Given a bounded open cube $\Lambda\subset\rz^{d}$, the
self-adjoint negative Dirichlet Laplacian is the operator $-\Delta_{\Lambda}:
\mathcal{D}(-\Delta_{\Lambda}) \ni\varphi\mapsto - \nabla^{2}\varphi$
with domain $\mathcal{D}(-\Delta_{\Lambda}) := \big\{
\varphi\in\RL^{2}(\Lambda) : 
\varphi, (\nabla\varphi)_{1},\ldots,(\nabla\varphi)_{d}$ are absolutely
continuous on $\Lambda$, $\nabla^{2}\varphi\in\RL^{2}(\Lambda)$
and $\varphi(x)=0$ for all $x\in\partial\Lambda\big\}$.

\begin{definition}
  \label{1.2.4}
  A \emph{random potential} $V$ on $\rz^{d}$ is an
  $\rz^{d}$-homogeneous random field $V:
  \Omega\times\rz^{d}\rightarrow\rz$, $(\omega,x)\mapsto 
  V^{(\omega)}(x)$, on a complete probability space
  $(\Omega, \mathcal{A},\bbP)$ which is jointly measurable with respect
  to the product of the sigma-algebra $\mathcal{A}$ of event sets in
  $\Omega$ and the sigma-algebra $\mathcal{B}^{\, d}$ of Borel sets in
  $\rz^{d}$. Moreover, defining the constants $p(d):=2$ for $d\le 3$,
  respectively  
  $p(d) > d/2$ for $d\ge 4$, we assume the existence of two reals
  $p_{1}> p(d)$ and $p_{2}> p_{1}d/[2(p_{1}-p(d))]$ such that 
  \begin{equation}
    \label{1.2.4glg}
    \bbE\bigl\{ \norm[p_{1};\Lambda_{1}(0)]{V}^{p_{2}}\bigr\} < \infty\,.
  \end{equation}
  Here, $\bbE\{\cdot\}:= \int_{\Omega}\d\bbP(\omega)\,(\cdot)$ is the
  expectation associated with the probability measure $\bbP$.
\end{definition}

For later purpose we recall from \cite{Dou94} the following 

\begin{definition} \label{strongmix}
  Given a random potential $V$ on $\rz^{d}$ and a Borel subset
  $\Lambda\subset\rz^{d}$, let the local sigma-algebra
  $\mathcal{A}_{V}(\Lambda)$ be the  
  sub-sigma-algebra of events generated by the  set of
  random variables $\{V^{(\cdot)}(x) :\, x\in\Lambda \}$. 
  The \emph{strong mixing coefficient} of $V$ is defined by 
  \begin{align}\label{smc}
    \alpha_{V}(L,G) := \sup\bigl\{ \kappa_{V}(\Lambda,\Lambda'): 
    \Lambda,\Lambda'\subset\rz^{d}; \dist_{\infty} &
    (\Lambda,\Lambda')\ge L; \nonumber\\
    & |\Lambda|,|\Lambda'| \le G\bigr\},
  \end{align}
  where $L,G>0$ and $\kappa_{V}(\Lambda,\Lambda'):= \sup\bigl\{ |\bbP(A
  \cap A') - \bbP(A)\bbP(A')| : A\in\mathcal{A}_{V}(\Lambda),
  A\in\mathcal{A}_{V}(\Lambda')\bigr\} \le 1/4$ measures the
  stochastic dependence
  of the restrictions of $V$ to $\Lambda$ and $\Lambda'$, respectively.
\end{definition}

Now we give the precise definition of random
Schr\"odinger operators of type \eqref{formalH} and collect some of
their basic properties in  

\begin{proposition}
  \label{Hdef}
  Let $a: \rz^{d}\rightarrow\rz^{d}$ be a non-random, continuously
  differentiable 
  vector field with vanishing divergence, $\nabla\cdot a = 0$, and let 
  $V$ be a random potential on $\rz^{d}$ in the sense of
  Definition~\ref{1.2.4}. 
  Then,
  \begin{nummer}
  \item given a bounded open cube $\Lambda\subset\rz^{d}$, the associated
    \emph{finite-volume random Schr\"odinger operator} with Dirichlet
    boundary conditions
    \begin{equation} \label{diriop}
      H_{\Lambda}(V^{(\omega)}):
      \mathcal{D}(-\Delta_{\Lambda}) \ni \varphi\mapsto \frac{1}{2}\,
      (\i\nabla +a)^{2}\varphi + V^{(\omega)}\varphi
    \end{equation}
    is self-adjoint on $\RL^{2}(\Lambda)$ and its spectrum is purely
    discrete for $\bbP$-almost all
    $\omega\in\Omega$. Therefore the \emph{finite-volume integrated
      density of states} 
    \begin{equation}
      \label{findos}
      N_{\Lambda}^{(\omega)}(E) := 
      \biggl\{\, 
        \begin{array}{l}
          \text{number of eigenvalues of
            $H_{\Lambda}(V^{(\omega)})$, counting} \\
          \text{multiplicity, which are strictly smaller than $E$}
        \end{array}
      \biggr\}
    \end{equation}
    associated with $H_{\Lambda}(V^{(\omega)})$ is well-defined for
    $\bbP$-almost all $\omega\in\Omega$.
  \item \label{Hdefess} 
    the operator $\mathcal{C}_{0}^{\infty}(\rz^{d}) \ni \varphi\mapsto
    \frac{1}{2}\, (\i\nabla +a)^{2}\varphi + V^{(\omega)}\varphi$
    is essentially self-adjoint for $\bbP$-almost all
    $\omega\in\Omega$. Its self-adjoint closure $H(V^{(\omega)})$ on
    $\RL^{2}(\rz^{d})$ is
    called (\emph{the infinite-volume}) \emph{random Schr\"odinger operator}. 
  \item \label{Hdefmess} 
    the mappings $\omega\mapsto H_{\Lambda}(V^{(\omega)})$ and
    $\omega\mapsto H(V^{(\omega)})$ 
    are measurable. The same is true for the projection-valued
    spectral measures of $H_{\Lambda}$ and $H$ associated with the
    pure-point, the absolutely 
    continuous and the singular continuous component in the Lebesgue
    decomposition of their spectra.
  \end{nummer}
\end{proposition}

\begin{remark}
  We will primarily be interested in spatially constant magnetic-field 
  tensors $\partial a_{j}/\partial x_{k} - \partial a_{k}/\partial x_{j}$,
  $j,k =1,\ldots, d$, and
  have thus dispensed with formulating Proposition~\ref{Hdef} under weaker
  assumptions on the vector potential $a$. The interested reader
  may consult e.g.\ \cite{HiSt92}, \cite{BrHuLe00} and \cite{FiHu98}. 
  Since the proof of
  Proposition~\ref{Hdef} is a standard result for $a=0$ \cite{Kir89,CaLa90},
  we will mainly comment on the changes required for $a\neq 0$.
\end{remark}

\begin{proof}[Proof of Proposition \ref{Hdef}]
  \begin{nummer} 
  \item Since the bound \eqref{1.2.4glg} ensures $V^{(\omega)} \in
    \RL_{\mathrm{loc}}^{p_{1}}(\rz^{d}) \subseteq
    \RL_{\mathrm{loc}}^{p(d)}(\rz^{d})$ for $\bbP$-almost all
    $\omega\in\Omega$,  it follows that for these $\omega$'s the operator 
    $\i a\cdot\nabla + a^{2}/2 + V^{(\omega)}$ is an operator
    perturbation of $-\Delta_{\Lambda}$ with relative operator bound
    zero. Self-adjointness of $H_{\Lambda}(V^{(\omega)})$ on
    $\mathcal{D}(-\Delta_{\Lambda})$ is then guaranteed by the
    Kato-Rellich theorem. Since operator boundedness with bound zero
    implies form boundedness with bound zero, the discreteness of the  
    spectrum of $H_{\Lambda}(V^{(\omega)})$ follows from that of 
    $-\Delta_{\Lambda}$ and the min-max principle, see e.g.\ Sect.~7.2
    in \cite{Kir89}.
  \item It is shown in the proof of Prop.\ V.3.2 in \cite{CaLa90}
    that \eqref{1.2.4glg} implies the $\bbP$-almost sure existence of a
    decomposition $V=V_{1}+V_{2}$ with $V_{1}\in
    \RL^{p(d)}_{\text{unif, loc}}(\rz^{d})$, $V_{2}\in
    \RL^{2}_{\text{loc}}(\rz^{d})$ and $V_{2}(x) \ge - c x^{2}$ for
    Lebesgue-almost all $x\in\rz^{d}$ with
    some constant $c>0$. Here we say that a measurable function
    $\varphi: \rz^{d}\rightarrow\cz$ belongs to the space
    $\RL^{p}_{\text{unif, loc}}(\rz^{d})$, if $\sup_{y\in\rz^{d}}
    \norm[p;\Lambda_{1}(y)]{\varphi} < \infty$.
    The claim thus follows from Thm.~2.5 in \cite{HiSt92}, since 
    $\RL^{p(d)}_{\text{unif, loc}}(\rz^{d})$ is a subset of both
    $\RL^{2}_{\text{loc}}(\rz^{d})$ and the Kato class over $\rz^{d}$,
    see e.g.\ Prop.~4.3 in \cite{AiSi82}.
  \item This is a consequence of the considerations in Sect.~V.1 of
    \cite{CaLa90} and of a straightforward generalization to non-zero
    vector potentials $a$ of Prop.~V.3.1 in \cite{CaLa90}. \qed
  \end{nummer}
\end{proof}

In the next proposition we recall some important properties of ergodic
random Schr\"odinger operators.
For assertion~\itemref{constspec} to hold, it is essential that, as usual, the
pure-point spectrum of an operator is defined as the closure of the
set of its eigenvalues.

\begin{proposition} \label{Herg}
  In addition to the requirements of Proposition~\ref{Hdef} assume that
  the vector potential $a$ gives rise to a constant magnetic-field
  tensor and that $V$ is ergodic with respect to translations in
  $\rz^{d}$. Then
  \begin{nummer}
  \item \label{constspec}
    The spectrum of $H(V)$, as well as its components in the
    Lebesgue decomposition -- the pure-point spectrum, the absolutely
    continuous spectrum and the singular continuous spectrum -- are
    $\bbP$-almost surely equal to non-random closed subsets of the
    real line $\rz$.
  \item \label{dosbem}
    If, in addition, $\bbE\{\exp [-t V(0)]\} < \infty$ for all
    $t>0$, there exists a non-random left-continuous distribution 
    function $N$ on $\rz$, called the
    \emph{\emph{(}infinite-volume\emph{)} integrated 
      density of states}, such that 
    \begin{equation} \label{dos}
      N(E) = \lim_{\Lambda  \uparrow\rz^d} 
      \frac{N_{\Lambda}^{(\omega )}(E)}{|\Lambda |} \,.
    \end{equation}
    More precisely, there is a set $\Omega _{0}\in \mathcal{A}$ of
    full probability, $\bbP(\Omega _{0}) =1$, such 
    that (\ref{dos}) holds for all $ \omega\in\Omega _{0} $
    and for all $E\in \rz$ except for the at most countably many
    discontinuity points of $N$. 
  \item \label{pasrep}
    If, in addition, $\bbE\{|V(0)|^{r}\} <\infty$ for some $r \ge r'
    +1$, where $r'$ is the smallest even integer with $r'>d/2$,
    one has the representation 
    \begin{equation}
      \label{tracerep}
      N(E) = \norm{f}_{2}^{-2} \bbE\Bigl\{ \Tr\Bigl( f^{*}\, \varTheta
      \bigl( E - H(V)\bigr) f \Bigr)\Bigr\}
    \end{equation}
    with any non-zero $f\in \mathrm{L}^{2}(\rz^{d})$, which, inside
    the trace, is to be understood as an operator of multiplication.
    Moreover, $\varTheta\bigl( E - H(V^{(\omega)})\bigr)$
    denotes the spectral projection operator of $H(V^{(\omega)})$ associated
    with the open interval $]-\infty, E[$. As a consequence, the
    set of growth points of $N$ coincides with the $\bbP$-almost sure
    spectrum of $H(V)$.
  \end{nummer}
\end{proposition}

\begin{proof}
  Concerning assertion~\itemref{constspec} we refer to Thm.~1 in \cite{KiMa82}.
  The existence and non-randomness  of the integrated
  density of states is shown in Thm.~VI.1.1 in \cite{CaLa90} for the
  case $a=0$ by using 
  functional-integral techniques for the Laplace transforms of the
  density-of-states measures. Employing the appropriate
  Feynman-Kac-It\^o formula and so-called magnetic translations, these
  methods generalize in a 
  straightforward manner to the present setting with a constant
  magnetic field \cite{BrHuLe93,Uek94}. Note also that
  pointwise convergence of the  
  Laplace transforms of a sequence of measures implies pointwise
  convergence of the associated distribution functions at all
  continuity points of the limit. Alternatively, \eqref{dos} may be
  obtained from a purely functional-analytic argument, which is outlined in
  \cite{Mat93}. 
  The representation \eqref{tracerep} claimed in \itemref{pasrep} is
  contained in \cite{PaFi92} as Thm.~5.20 and Prob.~II.4 in the case
  $a=0$; for the extension to $a\neq 0$ see \cite{FiHu98}. The proof
  of \eqref{tracerep} uses the resolvent of $H(V)$. In contrast, the
  proof of part \itemref{dosbem} 
  relies on semigroup techniques. This explains the different
  assumptions in \itemref{dosbem} and \itemref{pasrep}. 
  The assertion on the growth points of $N$ follows from \eqref{tracerep}.
\end{proof}

\subsection{Gaussian Random Potentials and Main Result}
\label{resultsection} 

A random field on $\rz^{d}$ is called Gaussian, if all its
finite-dimensional marginal distributions are Gaussians. Such a random
field is completely characterized -- up to equality in distribution --
by its mean function $\rz^{d}\ni x\mapsto \bbE\{V(x)\}$ and its
covariance kernel $\rz^{d}\times\rz^{d} \ni (x,y)\mapsto \bbE\{ V(x)
V(y)\}$. The reader is referred to \cite{Adl81, Lif95, Yur95} for
detailed expositions about Gaussian random fields.

\begin{definition} \label{Gaussrp}
  A \emph{Gaussian random potential} on $\rz^{d}$ (with zero mean) is
  an $\rz^{d}$-homogeneous Gaussian random field $V:
  \Omega\times\rz^{d} \rightarrow \rz$, $(\omega,x) \mapsto
  V^{(\omega)}(x)$, on a complete probability space $(\Omega,
  \mathcal{A}, \bbP)$ such that $\bbE\{V\} = 0$ and the covariance
  function $\rz^{d}\ni x \mapsto C(x):=\bbE\{V(x)V(0)\}$ is continuous
  at the origin where it obeys $0<C(0)<\infty$.
\end{definition}

\begin{remark}  \label{ellce}
  Given a Gaussian random potential on $\rz^{d}$ there exists a length
  $\ell_{C}>0$ such that $C(x)>0$ for all $x\in\Lambda_{\ell_{C}}(0)$
  due to our continuity requirement.
\end{remark}

\begin{lemma}
  \label{gauss1.2.4}
  A Gaussian random potential on $\rz^{d}$ is a random potential on
  $\rz^{d}$ in the sense of Definition~\ref{1.2.4}.
\end{lemma}

\begin{proof}
  A covariance function $ C $ which is continuous at the origin, where
  it satisfies
  $C(0) < \infty $, is bounded and uniformly continuous on $ \rz^d $ by
  the Bochner-Khintchine theorem. Consequently, Thm.~3.2.2 in
  \cite{Fer75} implies 
  the existence of a separable version of $V$ which is jointly
  measurable with respect to the sigma-algebra $\calA$ and 
  the Borel sigma-algebra on $\rz^d$. From now on it is tacitly
  assumed that only this version will be dealt with when we refer to
  a Gaussian random potential. It remains to verify \eqref{1.2.4glg}.
  To this end choose an even natural number $p_{1} >
  p(d) + d/2$. Then there exists $p_{2}\in\rz$ such that $p_{1} > p_{2}
  > p_{1}d/[2(p_{1} - p(d))]$. Jensen's inequality, the homogeneity of
  $V$ and the explicit computation of the arising Gaussian integral
  now imply
  \begin{align}
    \bbE\bigl\{\norm[p_{1};\Lambda_{1}(0)]{V}^{p_{2}}\bigr\} 
    & \le \bigl(\bbE\bigl\{ \norm[p_{1};\Lambda_{1}(0)]{V}^{p_{1}}\bigr\} 
    \bigr)^{p_{2}/p_{1}} 
    =  \bigl(\bbE\bigl\{ |V(0)|^{p_{1}}\bigr\} \bigr)^{p_{2}/p_{1}}
    \nonumber\\ 
    & =  \bigl(C(0)\bigr)^{p_{2}/2} \prod_{k=1}^{p_{1}/2} (2k-1)^{p_{2}/p_{1}}
    < \infty\,. \qed
  \end{align}
\end{proof}

\begin{remark} \label{gaussapply}
  Given a Gaussian random potential $V$, then Lemma \ref{gauss1.2.4}
  allows to define the associated random Schr\"odinger operators  as
  in Proposition~\ref{Hdef}. Proposition~\ref{Herg} is then applicable,
  too, under the 
  additional assumptions stated there. Note also that the additional
  requirements in parts \itemref{dosbem} and \itemref{pasrep} of 
  Proposition~\ref{Herg} are fulfilled because of    
  $\bbE\{\exp [-t V(0)]\} = \exp\{t^{2}C(0)/2\} < \infty$ for all $t\in\rz$. 
\end{remark}

Before we formulate a spectral-localization theorem for Schr\"odinger
operators 
with Gaussian random potentials, it is convenient to define a couple
of properties which a Gaussian random potential on $\rz^{d}$ may have
or not.

\begin{indentnummer*}
\item[\ass{P}] Its covariance function $C$ is non-negative.
\item[\ass{H}] $C$ is locally H\"older continuous at the origin with
  some H\"older exponent $0< \beta \le 1$, that is,
  $C(0)-C(x) \le b \, |x|_{\infty}^{\beta }$ for all $x$ in some
  neighbourhood of  
  the origin and some constant $b > 0$.
\item[\ass{R}] $C$ admits the representation
  \begin{equation}\label{Frep}
    C(x)=\int_{\rz^d} \!\d^d y \; \gamma (x+y) \gamma (y) \, ,
  \end{equation}
  where $\gamma$ is a non-negative function on $\rz^{d}$  which satisfies
  the inequality $\gamma(x) \le \gamma_0 (1+|x|)^{-\zeta}$ for all
  $x\in\rz^{d}$ with some constants $\gamma_{0} >0$ and
  $\zeta >13\,d/2 +1$. Moreover, it is assumed to be uniformly
  H\"older continuous with some H\"older exponent $0< \alpha \le 1$,  
  that is, $|\gamma (x+y) - \gamma (x)| \le a \, |y|_{\infty}^{\alpha }$ 
  for all $x \in \rz^d$, all $y$ in some neighbourhood of 
  the origin and some constant $a>0$. 
\item[\ass{D}] $C$ decays (at least) algebraically at infinity,
  $|C(x)| \le C_{0} (1+|x|)^{-z}$ for all $x\in\rz^{d}$ with some constants
  $0<C_{0}<\infty$ and $z>4d + 3/2$. 
\item[\ass{E}] $V$ is ergodic.
\item[\ass{M}] There are constants $K_{0}\ge 2$ integer, $A >0$,
  $1<\nu < 1+ (8d)^{-1}$ and $\delta > 4(d-1)(\nu-1)/(2-\nu)$ such
  that the strong-mixing coefficient \eqref{smc} of $V$ satisfies the
  inequality
  \begin{equation}
    \label{smixbound}
    \alpha_{V} \bigl(l^{\nu}/4,(K_{0}-1)l^{d}\bigr) \le A (1 +l)^{-\delta}
  \end{equation}
  for all lengths $l >0$.
\end{indentnummer*}

\begin{remarks}
  \begin{nummer}
    \item \label{hoecont}
      The H\"older continuity property \ass{H} implies the
      $\bbP$-almost sure continuity, and hence local boundedness, of
      the realizations of $V$, see Section \ref{fluctSec}.
    \item The existence of the representation \eqref{Frep} in itself
      with some 
      $\gamma \in\RL^{2}(\rz^{d})$ is equivalent to the requirement
      that $C$ is the Fourier transform of a non-negative integrable
      function on $\rz^{d}$. This is in fact a rather weak
      requirement which is sometimes referred to as the
      Wiener-Khintchine criterion.
    \item If $C$ had a representation \eqref{Frep} with $\gamma$
      obeying $\int_{\rz^{d}}\d^{d}x\; \gamma(x) = 0$, then the
      Gaussian random potential could be constructed from a sequence
      of Poissonian random potentials in a suitably combined limit of
      infinite concentration of impurities and zero coupling of the
      single-impurity 
      potential. Within this interpretation $\gamma$ would appear as the
      scaled impurity potential of the Poissonian random potential.
    \item \label{Cabfall}
      Property \ass{D} implies property \ass{E}, because according
      to Example 1.15(c) in \cite{PaFi92} the decay of $C$ at infinity
      implies even mixing.  
    \item \label{zetaC}
      Property \ass{R} implies properties \ass{P}, \ass{H} and
      \ass{D}. If, in addition, $\gamma$ has compact support,
      then property 
      \ass{M} is implied, too. Indeed, for Gaussian random potentials
      $V$ the local sigma-algebras $\mathcal{A}_{V}(\Lambda)$ and
      $\mathcal{A}_{V}(\Lambda')$ are independent, if
      $\dist_{\infty}(\Lambda, \Lambda') > L$ with $L$ such that 
      the support of $C$ obeys
      $\mathrm{supp\,} C\subset \Lambda_{2L}(0)$.
    \item \label{Rimp}
      Property \ass{M} implies that the random potential $V$
      is strongly mixing. More precisely,
      $\lim_{L\to\infty} 
      L^{\mu}\alpha_{V}(L,G) =0$ for all $G>0$ and all $\mu <
      \delta/\nu$. The strong-mixing property is considerably
      stronger \cite{Dou94} than the usually required ergodicity
      \ass{E}. We do not 
      know however whether properties \ass{M} and \ass{D} are
      related to each other. In one dimension,
      strongly mixing random fields are completely non-deterministic
      (in other words: regular), see e.g.\ \cite{Gri81} or
      p.~21 in \cite{Dou94}.
      Whereas for one
      dimension there is a well-developed theory \cite{IbRo78}   
      characterizing the covariance functions of strongly mixing
      Gaussian random potentials, this seems to be an open problem for
      higher dimensions. However, more is known in the discrete case,
      that is, for random fields on the $d$-dimensional simple cubic
      lattice $\bbZ^d$, see e.g. Sect.~2.1 in \cite{Dou94}.
  \end{nummer}
\end{remarks}

\begin{examples}
  \begin{nummer}
  \item The Gaussian random potential on $\rz$ characterized
    by the exponential covariance $C(x)= C(0) \exp\{ -|x|/\xi\}$, $\xi
    >0$, has the 
    properties \ass{P}, \ass{H}, \ass{D} and \ass{M}. While the first
    three are
    obvious, the last one follows from Thm.~6 in Chap.~VI of \cite{IbRo78}.
  \item \label{detpot}
    The Gaussian covariance function $C(x) = C(0)\exp\{-x^{2}
    /(2\lambda^{2})\}$, $\lambda>0$, has the property \ass{R} for
    arbitrary dimension $d\ge 1$. Note that for $d=1$ this example
    gives rise to a so-called deterministic (in other words: singular)
    random field, whose realizations are 
    $\bbP$-almost surely real-analytic functions \cite{Bel59}. Hence,
    property \ass{M} does not hold in $d=1$, cf.\
    Remark~\ref{Rimp}. This is also meant as a warning that even a
    very fast decay of $C$ at infinity need not imply the
    strong-mixing property. 
  \end{nummer}
\end{examples}

Now we can state the \emph{main result} of this paper.

\begin{theorem}\label{maintheo}
  Let $V$ be a Gaussian random potential on $\rz^{d}$ with covariance
  function $\rz^{d} \ni x\mapsto \sigma^2 C(x)$, where $\sigma>0$ and
  $C$ has either 
  property \ass{R} or the four properties \ass{P}, \ass{H}, \ass{D}
  and \ass{M}. Let $H(V)$ be the associated random Schr\"odinger
  operator defined on the Hilbert space $\RL^{2}(\rz^{d})$ as in
  Proposition~\ref{Hdef}. 
  Then, given any $\sigma >0$ there is a finite energy  $E_{0} <0$
  such that the spectrum of $H(V)$ in the half-line $]-\infty, E_0]$ is
  $\bbP$-almost surely only pure point. 
  Moreover, given any finite energy $E_0 <0 $ there is a $\sigma_0 >0$
  such that the spectrum of $H(V)$ in $]-\infty, E_0]$ is
  $\bbP$-almost surely only pure point for all $\sigma \in \, ]0,\sigma_0]$.
\end{theorem}

\begin{remarks}
  \begin{nummer}
  \item If the vector potential $a$ generates a spatially constant
    magnetic-field tensor, then the assumptions of
    Theorem~\ref{maintheo} ensure that the components in the Lebesgue
    decomposition of the spectrum of $H(V)$ are $\bbP$-almost surely
    non-random sets, see Remark~\ref{gaussapply} and
    Proposition~\ref{constspec}. Moreover, the   
    spectrum of $H(V)$ is $\bbP$-almost surely equal to the whole real 
    line, as can be seen by following the steps in the proof of
    Thm.~5.34(i) in \cite{PaFi92}. 
  \item We believe that the assumption $C\ge 0$ is only a technical
    one. We need it for the proof of the Wegner estimate and in order
    to exclude the existence of the singular continuous spectrum. 
  \item The assumptions of the theorem require a compromise between local 
    dependence and global independence of $V$, as can be inferred from
    Remarks~\ref{hoecont}, \itemref{Cabfall}, \itemref{zetaC} and
    \itemref{Rimp}. 
    From a physical point of view, both requirements are plausible:
    The effective one-particle interaction potential
    should be smooth due to screening. By the same token it is expected
    that impurities do not influence each other over large distances.
  \item The assumptions of the theorem allow for deterministic random 
    potentials, cf.\ Example~\ref{detpot}. Since this is not the case
    for many other localization results for multi-dimensional space, the
    method of proof which we present here may also be interesting
    from a conceptual point of view.
  \item We are not able to prove exponential decay of the
    eigenfunctions $\psi_{n}^{(\omega)} \in \RL^2(\rz^d)$ associated with
    eigenenergies $E_{n}^{(\omega)}\in ]-\infty, E_{0}]$. If such a
    result is true, 
    it can certainly not be proven by a straightforward modification
    of our approach.
  \item \label{leider}
    It would be interesting to prove spectral localization also for
    strong disorder or at positive energies, e.g.\ in between Landau
    levels. Unfortunately, we do not know of the appropriate initial
    estimates for the multi-scale analysis.
  \item The assertions of the theorem remain true with $E_{0}$ below
    the bottom of the spectrum of $H(0)$, if $H(0)$ is generalized to
    include a (sufficiently well-behaved) periodic potential. For
    energies in the spectral gaps of $H(0)$ however, it is again the
    lack of appropriate initial estimates which prevents us from
    proving localization.
  \end{nummer}
\end{remarks}

Our proof of Theorem \ref{maintheo} relies on a multi-scale
analysis which requires a so-called Wegner estimate as an important
ingredient. Such an estimate bounds the mean number of eigenvalues in
a given energy interval of the finite-volume random Schr\"odinger operator 
by the length of the interval and the volume of the cube.

\subsection{Wegner Estimate for Gaussian Random Potentials} \label{wegnerSec}

In Theorem 1 and Remark 3 (ii) of \cite{FiHu97b} a Wegner estimate is 
obtained for Schr\"odinger operators with certain Gaussian random potentials
for the case without magnetic field. It was however already pointed
out there in the ``Note added in proof''
that the result continues to hold if a suitable magnetic vector potential is
included. Indeed, the two main technical steps in the proof remain
valid in the presence of a (continuously differentiable)
vector potential $a$. The first step
concerns the lowering of the eigenvalues of $H_{\Lambda}(V)$ in
inequality (27) of \cite{FiHu97b} by Dirichlet-Neumann bracketing and the
subsequent introduction of ``Neumann interfaces''. The second step is
the application of the abstract one-parameter spectral-averaging estimate
Cor.~4.2 of 
\cite{CoHi94}. Moreover, by applying the diamagnetic inequality for
Neumann partition functions, the Wegner constant \eqref{wegnerkonst}
below, derived for the case $a=0$ in \cite{FiHu97b}, is seen to be a Wegner
constant also for $a\neq 0$. For the technical details we refer to
\cite{FiHu98}.
We state the Wegner estimate as

\begin{proposition} \label{wegnerThm}
  Let $V$ be a Gaussian random potential on $\rz^{d}$ with property
  \ass{P}, let the associated finite-volume random Schr\"odinger
  operator be defined as in Proposition~\ref{Hdef} and define the
  length $\ell_{C}>0$ as in Remark~\ref{ellce}. 
  Then, for every energy $E\in \rz$ there is a 
  constant $0< W(E) <\infty$, depending on $\ell_{C}$,
  such that for all $E_{1},E_{2} \le E$ and
  all bounded open cubes $\Lambda\subset\rz^{d}$ with $|\Lambda| \ge
  \ell_{C}^{d}$ the averaged finite-volume integrated
  density of states \eqref{findos} obeys
  \begin{equation} 
    \label{wegner} 
    \bbE\left\{ \left| N_{\Lambda }(E_{1}) - N_{\Lambda }(E_{2}) \right|
    \right\} \le |\Lambda |\, |E_{1}-E_{2}| \, W(E)\,.
  \end{equation}
\end{proposition}

\begin{remarks}
  \begin{nummer}
  \item 
    The Wegner estimate \eqref{wegner} holds under weaker assumptions
    on $V$ which are stated in \cite{FiHu97b} or \cite{FiHu98}. Since
    we need stronger assumptions on $V$ in this paper in order to
    prove localization, we contented ourselves to formulate
    Proposition~\ref{wegnerThm} within the latter setting. 
  \item
    The Wegner constant $W(E)$ may be taken \cite{FiHu97b} as
    \begin{equation} \label{wegnerkonst}
      W(E) = \frac{\exp\{t E + t ^2 C_{E}/2\}}%
      {\sqrt{2\pi C(0)}\; b_{E}}\,\Big( 2 \ell_{E}^{-1} + 
      (2\pi t )^{-1/2}\Big)^d\,,
    \end{equation}
    where $t > 0$ is arbitrary and may be considered as a variational 
    parameter. In (\ref{wegnerkonst}) we are using the constants
    $\ell_{E} := \inf\{ |E|^{-1/2}, \ell_{C} \}$, $b_{E} :=
    \inf_{|x|_{\infty } < \ell_{E}/2} \{C(x)/C(0)\} >  0$ and 
    $ C_{E} := C(0) (2 - b_{E}^2)$.
    The simple (but not optimal) choice $t = (2 C_{E})^{-1} (-E +
    \sqrt{E^2 + 2C_{E}/\pi })$ gives
    the leading asymptotic low- and high-energy behaviour 
    \begin{equation} \label{wasym}
      \lim_{E \to - \infty}
      \frac{\ln W(E)}{E^2}=- \, \frac{1}{2\, C(0)}\,,
      \qquad\quad
      \lim_{E \to \infty}
      \frac{W(E)}{E^{d/2}}= 
      \frac{3^{d}\, \e^{1/(2\pi)}}{\sqrt{2\pi C(0)}}\,.
    \end{equation}
  \item
    The Lipschitz continuity (\ref{wegner}) of the averaged  finite-volume
    integrated density of states implies by the Chebyshev-Markov inequality 
    that the probability of finding the spectrum of $H_{\Lambda }(V)$
    near a given energy $\tilde{E}$ is controlled by the inequality
    \begin{equation} \label{wegnerloc}
      \bbP\left\{ \omega :
        {\rm dist}\Big( {\rm spec} \big(H_{\Lambda } 
        (V^{(\omega)})\big), \tilde{E} \Big)
        < \varepsilon \right\} 
      \le  2 |\Lambda | \, \varepsilon\, W(E ) \, .
    \end{equation}
    It is valid for all bounded open cubes 
    $\Lambda \subset \rz^d$ with $|\Lambda | \ge \ell_{C}^{d}$ 
    and for all energies $\tilde{E}\in \rz$ and $\varepsilon > 0$ such that 
    $\tilde{E} + \varepsilon  \le E$.
  \item Apart from \cite{FiHu97b}, we know of
    \cite{CoHi94, Kir96, KiSt98a,CoHi98,Zen99,Sto00}, where a Wegner
    estimate is proven for 
    correlated random potentials in multi-dimensional continuous space.
  \end{nummer}
\end{remarks}
                                
Recalling Remark~\ref{gaussapply} and Proposition~\ref{dosbem} on the
existence of the integrated density of states, we
conclude from the reasoning in \cite{FiHu97b} that
Proposition~\ref{wegnerThm} has the following  

\begin{corollary} \label{abscontCor}
  Assume that the vector potential $a$ generates a constant
  magnetic-field tensor and that the Gaussian random potential $V$ has
  the properties \ass{P} and \ass{E}.
  Then, the integrated density of states $N$ is
  absolutely continuous on any bounded interval and its derivative, 
  the density of states, is locally bounded in the sense that
  \begin{equation} \label{abscont}
    0\le \frac{\d N(E)}{\d E} \le W(E)
  \end{equation}
  for Lebesgue-almost all $E\in\rz$.
\end{corollary}

\begin{remarks}
  \begin{nummer}
  \item
    Of course, the assumption of a constant magnetic-field tensor is
    not really necessary. Corollary~\ref{abscontCor} holds for all vector
    potentials $a$ which give rise to an ergodic random Schr\"odinger
    operator $H(V)$. 
  \item
    One must not expect that \eqref{abscont} with $W(E)$ given by 
    \eqref{wegnerkonst} is a sharp bound 
    on the density of states $\d N/ \d E$, since neither special
    properties of the covariance function $C$ have entered nor has 
    the vector potential $a$. 
    Whereas we conjecture that \eqref{wasym} reflects the true asymptotic
    behaviour at low energies of the density of states $\d N/ \d E$, 
    this should
    not be true for the high-energy asymptotics. For example, we conjecture
    that the density of states behaves like $E^{d/2-1}$ for $E\to\infty$
    in the case $a=0$.
  \end{nummer}
\end{remarks}


\section{Multi-Scale Analysis} \label{multiscale}

The heart of the localization proof given in this paper is a
multi-scale analysis in the spirit of the fundamental work
\cite{FrSp83}. It provides probabilistic bounds on the resolvents of
finite-volume random Schr\"odinger operators. The technical
realization of the multi-scale analysis used here is patterned after
\cite{DrKl91} and \cite{CoHi94} in order to cope with a correlated
random potential and a continuous space, respectively; for the latter
aspect see also \cite{FiKl96}. In addition, to
overcome difficulties arising from long-range spatial correlations, we
allow for different random potentials on different length
scales. The idea behind this is to replace a given long-range
correlated random potential $V$ on the length scale $L_{k}$ by the
element $V_{k}$ of a sequence $\{V_{k}\}_{k\in\bbN_{\,0}}$ of random
potentials such that ~(i)~ $\{V_{k}\}$ converges to $V$ in a suitable
sense and ~(ii)~ each $V_{k}$ has short-range correlations, but its spatial
extent grows with increasing $k$.

We do not restrict ourselves to Gaussian random potentials in this
section, but only require the setting of Proposition~\ref{Hdef}. 
The main technical result of this section is
stated as Theorem~\ref{multThm} in Subsection~\ref{finvol}. In
Subsection~\ref{infvol} we perform the macroscopic limit to infinite
volume.

%
\subsection{Boxes and Geometric Resolvent Equation}
%

A \emph{box} $ B \subseteq \rz^d \times \rz_{\,0}^{+} $ is a pair
$(\Lambda ,b )$ consisting of an open cube $ \Lambda \subseteq \rz^d $ and
the width $ b\ge0 $ of the \emph{frame}
$ \{ x\in \Lambda : \dist_{\infty}(x,\partial \Lambda) \le b \} $ of $B$.
The box $ B=(\Lambda ,b ) $ is said to be contained in the
box $ B\, '=(\Lambda' ,b') $, in symbols, $ B < B\, ' $, if
$ \Lambda\subset\Lambda' $ and
$ \dist_{\infty}(\Lambda,\partial\Lambda') > b' $.

\begin{figure}[t]
  \begin{center}
    \hspace*{1.5cm}\parbox{12.8cm}{\setlength{\unitlength}{1.6cm}
%
\begin{picture}(8,4)
\put(0,1){\vector(1,0){6}} 
\put(6,0.7){$ x_{1}$}
\put(1,1){\vector(0,1){1.5}} 
\put(1,1){\circle{0.1}} 
%
%
\put(0.9,2){\line(1,0){0.2}}
\put(0.7,2.1){$1$}
\put(2.8,0.9){\line(0,1){0.2}}
\put(2.5,0.5){$ \frac{\ell}{2}-b $}
\put(3.6,0.9){\line(0,1){0.2}}
\put(3.3,0.5){$ \frac{\ell}{2}-\frac{2b}{3} $}
\put(4.4,0.9){\line(0,1){0.2}}
\put(4.2,0.5){$ \frac{\ell}{2}-\frac{b}{3} $}
\put(5.2,0.9){\line(0,1){0.2}}
\put(5.2,0.5){$\frac{\ell}{2}$}
%
%
\bezier{200}(2.8,1)(3,1)(3.2,1.5)
\bezier{200}(3.2,1.5)(3.4,2)(3.6,2)
\put(3.6,2){\line(1,0){0.8}}
\bezier{200}(4.4,2)(4.6,2)(4.8,1.5)
\bezier{200}(4.8,1.5)(5,1)(5.2,1)
\put(5.1,1.8){$\indfkt{B}^{\partial}(x)$}
\linethickness{0.7pt}
\bezier{70}(0.4,2)(2,2)(3.6,2)
\bezier{13}(3.6,2)(3.8,2)(4,1.5)
\bezier{13}(4,1.5)(4.2,1)(4.4,1)
\put(2,2.2){$\indfkt{B}(x)$}
\end{picture}}
    \caption{\label{indfunktfig}
       Sketch of indicator functions of the box
       $B=(\Lambda_l(0), b)$ (dotted line) and of the frame of $B$
       (solid line) as a function of $x_1$ for $x_j=0$,
       $j=2,\ldots,d$.}
  \end{center}
\end{figure}
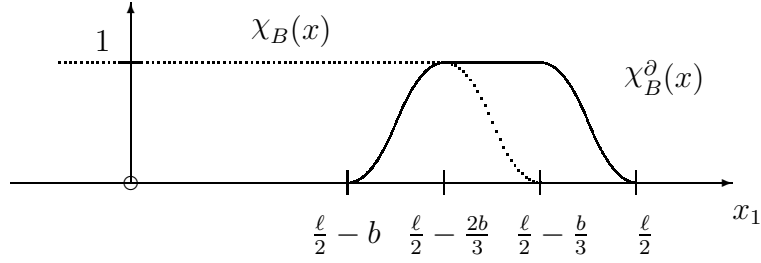

By definition, a (\emph{mollified\emph{)} indicator function}
$\indfkt{B}$ of a box 
$ B:=(\Lambda_\ell(y),b)  $ $ \subset \rz^d\times \rz_{\,0}^{+} $
has the properties
\begin{indentnummer}
\item $ 0 \le \indfkt{B} \le 1 $,
\item $\indfkt{B}(x) =
  \begin{cases}
    1 & \text{if } |x-y|_{\infty} < \ell/2-2b/3 \, ,\\
    0 & \text{if } |x-y|_{\infty} \ge \ell/2-b/3 \, .
  \end{cases}
  $
\end{indentnummer}
Moreover, if $ b>0 $ we infer from Uryson's lemma in the version of 
Thm.~1 of Sect.~3.4 in \cite{Ric78} that
\begin{indentnummer} \setcounter{numcount}{2}
\item  $\indfkt{B} \in \calC^\infty_0(\rz^d)$, 
\item
  there exist positive real constants $ \varkappa_1 , \varkappa_2,
  \varkappa_4 >0 $, which are independent of $ B $, such that
  \begin{align}
    \norm[\infty]{\nabla\indfkt{B}} & \le \varkappa_1 /b\,,
    \label{kappa1}\displaybreak[0]\\[1mm]
    \norm[\infty]{\nabla^{2}\indfkt{B}} & \le \varkappa_2 /b^{2}\,,
    \label{kappa2}\displaybreak[0]\\[1mm]
    \norm[\infty]{\nabla^{2}|\nabla \indfkt{B}|^{2}} & \le
    \varkappa_4/b^{4}\,.
    \label{kappa4}
  \end{align}
\end{indentnummer}
We set $\indfkt{(\rz^{d},0)}:=1 $.
For $b>0$ we say that $\indfkt{B}^{\partial}$ is an  
\emph{indicator function of the frame} of the box $B$, if it has 
the properties
\begin{nummer}
\item
  $ \indfkt{B}^{\partial} \in \calC^\infty_0(\rz^d) $,
\item
  $ 0 \le \indfkt{B}^{\partial} \le 1 $, 
\item
  $\indfkt{B}^{\partial}(x)=
  \begin{cases}
    1 & \text{if } \ell/2 -2b/3 \le |x-y|_{\infty} \le \ell/2-b/3 \, ,\\
    0 & \text{if } |x-y|_{\infty} \le \ell/2-b \text{ ~or~ }
    |x-y|_{\infty} \ge \ell/2\,. 
  \end{cases}
  $
\end{nummer}
For an illustration of $\indfkt{B}$ and $\indfkt{B}^{\partial}$ see 
Fig.~\ref{indfunktfig}. 
Finally, we introduce
a \emph{frame operator} $ \varGamma_{B} $ as
the first-order differential operator given by
\begin{equation} \label{frameop}
\varGamma_{B}:= \frac{1}{2} \; ( \nabla^{2} \indfkt{B}) -
( \Ri \nabla \indfkt{B}) \cdot (\Ri \nabla + a )
\end{equation}
with domain
$ {\calD}(-\Delta _{\Lambda}) \subset \RL^2(\Lambda)$. Here
$ \nabla^{n} \indfkt{B} $, $ n= 1,2 $, acts as a multiplication operator,
and $a$ is the vector potential, see Proposition~\ref{Hdef}.
\begin{remark}
Obviously, the following operator identities hold
\begin{equation} \label{machtnix}
\indfkt{B}^{\partial}\, \nabla^{n} \indfkt{B}=
\nabla^{n}\indfkt{B} \, , \quad n=1,2\,,
\end{equation}
\begin{equation}
\varGamma_{B}= \indfkt{B}^{\partial}\, \varGamma_{B}=
\varGamma_{B}\, \indfkt{B}^{\partial} \qquad \text{on \,} {\cal D}
(-\Delta_\Lambda) \text{ \,for \,} B=(\Lambda ,b)\,.
\end{equation}
\end{remark}
For the rest of this section let $v\in\RL^{p(d)}_{\text{loc}}(\rz^d)$  
be a non-random real scalar potential, let $a:
\rz^{d}\rightarrow\rz^{d}$ be a continuously differentiable vector
potential with $\nabla\cdot a=0$
and let $H_{\Lambda}(v)$ be
defined as in \eqref{diriop}. The constant $p(d)$ was defined
in Definition~\ref{1.2.4}.
\begin{lemma}\label{geomreslemma}
For $b, b' >0 $ let $ B=(\Lambda ,b ) $ and
$ B\, '=(\Lambda' ,b') $ be boxes such that
$ B < B\, ' $. Then for all $ z\in \bbC \setminus \rz $ the geometric
resolvent equation
\begin{multline} \label{geores}
\big(H_{ \Lambda' }(v)-z\big)^{-1} \indfkt{ B }= \\
\indfkt{ B } \big(H_{ \Lambda }(v)-z\big)^{-1} +
\big(H_{ \Lambda' }(v)-z\big)^{-1} \varGamma_{ B }
\big(H_{ \Lambda }(v)-z\big)^{-1}
\end{multline}
holds on $ \RL^{2}(\Lambda') $. 
\end{lemma}

\begin{remark}
\begin{nummer}
\item For (\ref{geores}) to make sense as an operator identity on 
$ \RL^{2}(\Lambda') $, operators which are \emph{a priori} only 
defined on $ \RL^{2}(\Lambda) $ are supposed to be trivially extended to
$ \RL^{2}(\Lambda') $ by setting them to zero on 
$ \RL^{2}(\Lambda'\setminus \Lambda) $. This convention will be used 
throughout the paper without further notice.
\item The proof of Lemma \ref{geomreslemma} follows from writing
$ \varGamma_B = \indfkt{B} \bigl( H_\Lambda (v) -z\bigr) - 
\bigl( H_\Lambda (v) -z\bigr) \indfkt{B}$.   
\end{nummer}
\end{remark}

In what follows it will be useful to introduce the abbreviations
\begin{equation}
R_{ B\,' ,\, B }(v,z) :=\varGamma_{ B\, ' }
\big(H_{ \Lambda '}(v)-z\big)^{-1}\indfkt{ B }
\end{equation}
and
\begin{equation}
W_{ B\,' ,\, B }(v,z) :=\varGamma_{ B\, ' }
\big(H_{ \Lambda '}(v)-z\big)^{-1}\indfkt{ B }^{\partial}\, .
\end{equation}
We will omit the arguments $ v $ and $ z $ when this does not
cause confusion.
\begin{remarks}\label{RWRems}
\begin{nummer}
\item
The definitions of $ \indfkt{  } $, $ \indfkt{   }^{\partial} $,
$ R $ and $ W $ imply for all $ B < B\,' < B\,'' $ the relations
\begin{align}\label{RCube}
R_{ B\,'', B\,' } \,\indfkt{B } & = R_{ B\,'' ,B }\; , \\[1mm]
R_{ B\,'', B\,' } \,\indfkt{B }^{\partial} & = W_{ B\,'' ,B }\; .
\label{RFrame}
\end{align}
\item
Using the definitions of $ R $ and $ W $, the geometric resolvent
equation of Lemma \ref{geomreslemma} with boxes $ B<B\,'$ implies
\begin{equation}\label{RGeomRes}
R_{ B\,',B } (v,z) =W_{ B\,',B } (v,z)\, R_{B,(\Lambda,0)} (v,z)\; .
\end{equation}
\end{nummer}
\end{remarks}

We finish this subsection with a lemma which estimates $ R_{B',B} $ in
terms of a ``localized'' resolvent by bounding $\varGamma_{B'}$. This result
will only be needed in Section~\ref{applSec}.
\begin{lemma}\label{boundFrame}
Consider two boxes $ B =(\Lambda ,0) $ and $ B' =(\Lambda' ,b') $
with $ b'>0 $, $\Lambda \subseteq \Lambda'$ and $\Lambda'$
bounded. Then one has 
\begin{multline}
\sup_{ \eta > 0 }
\norm{R_{B',B}(v,E+ \Ri \eta) } 
\le \varPhi_{B'} (E-v) \;
\sup_{ \eta > 0 }\,
\bignorm{\indfkt{B'}^{\partial} 
\big( H_{\Lambda'}(v) -E- \Ri \eta \big)^{-1}
\indfkt{B} } \\
+
\varTheta \big( b'-\dist_{\infty}(\partial \Lambda', \Lambda) \big)
\, \frac{2^{1/2}\varkappa_1}{b'}\;
\sup_{ \eta > 0 } \,
\bignorm {\big( H_{\Lambda'}(v) -E- \Ri \eta \big)^{-1} }^{1/2}
\end{multline}
for Lebesgue-almost all $E \in \rz$. Here we 
used Heaviside's unit-step
function $ \varTheta (t) := \biggl\{ \begin{array}{r@{\;\text{ if }}l}
1 & t > 0 \\ 0 & t \le 0 \end{array} $ and  
introduced the functional
\begin{equation} \label{gfunctional}
\varPhi_{B'}(f) := \frac{1}{{b'}^{2}}\,
\left(
\frac{\varkappa_2}{2} \, + 
\sqrt{ \frac{\varkappa_4}{2} \, + 2\,(b'\varkappa_1)^2
\norm[\infty;\Lambda']{\max \{ 0, f \} } } \,
\right) \,,
\end{equation}
which is defined for real-valued functions 
$ f\in\RL^{\infty}(\Lambda')$. If $ f\notin\RL^{\infty}(\Lambda')$,
we set $\varPhi_{B'}(f):=\infty$.
\end{lemma}

The proof of Lemma \ref{boundFrame} relies on a partial-integration result
for the scalar product \eqref{scalarproduct} on  
$\RL^2(\Lambda)$.
It is an extension to non-zero vector potentials $a$ of Eq.~(2.7) 
in \cite{CyFr87}  
and replaces the wrong equality at the bottom of p.~175 in \cite{CoHi94}.

\begin{lemma} \label{pilemma}
  Let $\Lambda$ be a bounded open subset of $\rz^{d}$, let
  $\psi\in\mathcal{D}(-\Delta_{\Lambda})$, let
  $\phi\in\mathcal{C}^{2}(\rz^{d})$ real and let
  $a:\rz^{d}\rightarrow\rz^{d}$ be continuously differentiable with
  $\nabla\cdot a =0$, then
  \begin{equation}
    \label{pieq}
    \langle\phi,|(\i\nabla +a)\psi|^{2}\rangle =
    \frac{1}{2}\,\langle\nabla^{2}\phi,|\psi|^{2}\rangle
    +  \Re \bigl\{\langle\psi\phi, (\i\nabla +a)^{2}\psi
    \rangle\bigr\}\,.
  \end{equation}
\end{lemma}

\begin{proof}
  Using $|\nabla\psi|^{2} = (1/2)\nabla^{2} |\psi|^{2}  -  \Re\{ \psi^{*}
  \nabla^{2}\psi\}$, we rewrite the left-hand side of \eqref{pieq} as
  \begin{equation}
    \Bigl\langle \phi, \Bigl((1/2)\nabla^{2}|\psi|^{2} + 
    \Re\{\psi^{*}(\i\nabla)^{2}\psi\} + a^{2} |\psi|^{2} + 
    2\Re\{(\i\nabla\psi)\cdot a\psi^{*}\}\Bigr)\Bigr\rangle\,.
  \end{equation}
  The first term on the right-hand side of \eqref{pieq} follows from a
  double partial 
  integration and the rest from the reality of $\phi$ and the gauge
  condition $\nabla\cdot a=0$.
\end{proof}

\begin{proof}[Proof of Lemma \ref{boundFrame}]
  We infer from the definition \eqref{frameop} of $\varGamma_{B'}$
  and \eqref{kappa2} that
  \begin{equation}
    \label{bframestart}
    \bignorm{\varGamma_{B'}\psi} \le \frac{\varkappa_{2}}{2{b'}^{2}}\,
    \bignorm{\indfkt{B'}^{\partial}\psi} +
    \bignorm{(\nabla\indfkt{B'})\cdot(\i\nabla + a)\psi}\,.
  \end{equation}
  When applied to $\psi = \bigl( H_{\Lambda'}(v)-E-\i\eta\bigr)^{-1}
  \indfkt{B}\,g$, where $g\in\RL^{2}(\Lambda')$ with $\norm{g}=1$, Eq.\
  \eqref{bframestart} yields the claim of the lemma, provided the
  second term on the right-hand side of \eqref{bframestart} is shown
  to be appropriately bounded in terms of
  $\bignorm{\indfkt{B'}^{\partial}\psi}$. To do so we estimate
  \begin{align}
    \bignorm{(\nabla\indfkt{B'})\cdot(\i\nabla + a)\psi}^{2} &\le
    \bignorm{\,|\nabla\indfkt{B'}|\,|(\i\nabla + a)\psi|\,}^{2}
    \nonumber\\
    &=
    \langle |\nabla\indfkt{B'}|^{2}, |(\i\nabla + a)\psi|^{2}\rangle\,.
  \end{align}
  Since
  $\mathcal{D}(H_{\Lambda'}(v))=\mathcal{D}(-\Delta_{\Lambda'})$, we
  can apply Lemma~\ref{pilemma} to the scalar product. Together with
  \begin{equation}
    (\i\nabla +a)^{2}\psi = 2\indfkt{B}\,g + 2 (E + \i\eta - v)\psi
  \end{equation}
  this gives
  \begin{eqnarray}
    \lefteqn{
      \bignorm{(\nabla\indfkt{B'})\cdot(\i\nabla + a)\psi}^{2} 
      }\nonumber\\
    &&\le \frac{1}{2}\,\langle \nabla^{2} |\nabla\indfkt{B'}|^{2}, 
    |\psi|^{2}\rangle + 2 \Re\bigl\{\bigl\langle
    \psi|\nabla\indfkt{B'}|^{2}, \bigl(\indfkt{B}\,g + (E+\i\eta - v)
    \psi\bigr)\bigr\rangle\bigr\} \nonumber\\ 
    &&\le \left( \frac{\varkappa_{4}}{2{b'}^{4}} \, + 
    \frac{2\varkappa_{1}^{2}}{{b'}^{2}} \,  \norm[\infty;\Lambda']{\max\{0,
    E-v\}}\right)\;
    \bignorm{\indfkt{B'}^{\partial}\psi}^{2}\nonumber\\
    && \phantom{\le} + \varTheta \big( b'-\dist_{\infty}(\partial
    \Lambda', \Lambda) \big)\,\frac{2\varkappa_1^{2}}{{b'}^{2}}\;
    \norm{\psi}\,. \label{last3.5}
  \end{eqnarray}
  To derive the first line of the last inequality, we refer to
  \eqref{machtnix}, \eqref{kappa4} and \eqref{kappa1}. The second line
  follows from the Schwarz inequality, \eqref{kappa1} and
  $\norm{g}=1$. Thus, the claim is obtained by inserting
  \eqref{last3.5} into \eqref{bframestart} and by observing
  $\sqrt{\alpha^{2} + \beta^{2}} \le \alpha + \beta$ for
  $\alpha,\beta \ge 0$. 
\end{proof}

%
\subsection{Estimating Resolvents on Multiple Length Scales by Induction}
\label{finvol}
\begin{figure}[t]
\begin{center}
\parbox{10cm}{%
    \input{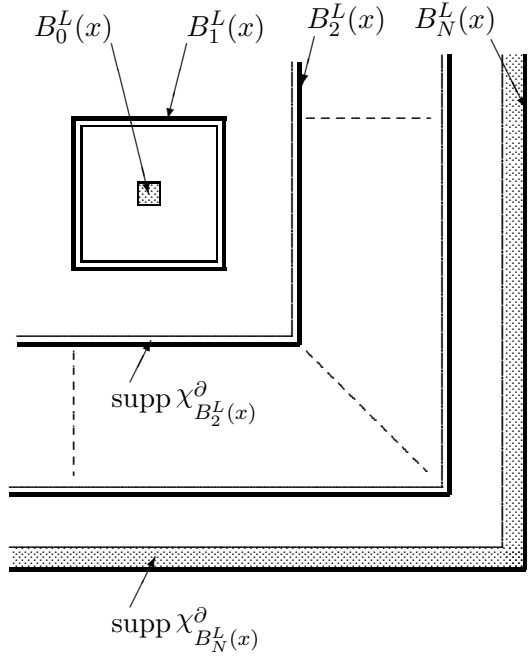}
}
\vspace{3ex}
\caption{\label{box1} 
Sketch for $d=2$ of the boxes defined in (\ref{boxDef1}).}
\end{center}
\end{figure}

%
%
Throughout this subsection we will consider a fixed real number $\nu>1$
and a fixed natural number $N>4$. Given a length $L>2^{2\nu/(\nu-1)}N$ and
a point $x\in \rz^{d}$, we define a family of $N+1$ nested boxes
\begin{equation}\label{boxDef1}
B^{L}_{n}(x) :=
 \begin{cases}  
   \;\Big(\Lambda_{L/(4N)}(x),0\Big)& \text{ if } n=0 \; ,\\[1.5ex]
   \;\left(\Lambda_{nL/N}(x),\frac{(L/N)^{1/\nu}}{4N}\right)
   & \text{ if } 1\le n \le N-1 \; ,\\[1.5ex]
   \;\Big(\Lambda_L(x),L/(4N)\Big) & \text{ if } n=N \; ,
 \end{cases}
\end{equation}
see Fig.\ \ref{box1}.
\begin{definition}
Let $E\in\rz$ and $r>0$. The potential $ v $ is said to be 
$ (r,E,L,x) $-\emph{regular}, if
\begin{equation}
\sup_{\eta>0}
\bignorm{R_{B_{N}^{L}(x),B_{0}^{L}(x)}(v,E+\i\eta)} \le L^{-r}\,.
\end{equation}
\end{definition}

\begin{figure}[t]
\begin{center}
\parbox{10cm}{%
    \input{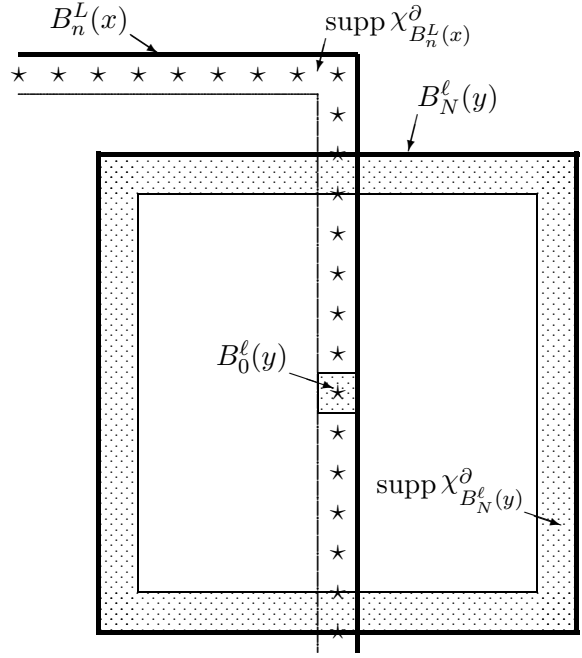}
}
\vspace{3ex}
\caption{\label{box2} 
Sketch of the geometric situation on the frame of $B_n^{L}(x)$ for
$1\le n \le N-1$ and $d=2$. The points marked by $\star$ make up the
set $T_n^{L}(x)$ and are the centres of the ``tiling'' of the frame of
$B_n^{L}(x)$. For one $y\in T_n^{L}(x) $ the boxes $B_0^{\ell}(y)$ and
$B_N^{\ell}(y)$ are sketched.}
\end{center}
\end{figure}

Now we come to the deterministic part of the recursion clause of the
multi-scale analysis. The aim is to infer the
$(r,E,L,x)$-regularity of a suitable potential $v'$ from the 
$(r,E,\ell,y)$-regularity of $v$
on the smaller length $\ell:=(L/N)^{1/\nu}$ for suitable $y\in\rz^{d}$.
To this end, choose a natural number $2\le S\le N-1$, a sequence
$\{n_s \}_{s=1, \ldots , S}$ of natural numbers with
$1 \le n_1 < \ldots < n_S \le N-1$, and apply (\ref{RCube}) and
(\ref{RGeomRes}) to get
$ R_{B_{N}^{L}(x),B_{0}^{L}(x)} =
W_{B_{N}^{L}(x),B_{n_1}^{L}(x)} R_{B_{n_1}^{L}(x),B_{0}^{L}(x)} $, where
the pair of arguments $(v,E+\i\eta)$ has been suppressed in the three 
operators.
Using (\ref{RFrame}) and (\ref{RGeomRes}) we derive the relation
$ W_{B_{N}^{L}(x),B_{n_s}^{L}(x)}=
W_{B_{N}^{L}(x),B_{n_s +1}^{L}(x)} W_{B_{n_s +1}^{L}(x),B_{n_s}^{L}(x)}$
for $s=1, \ldots , S-1 $ which, upon iteration with respect to  $s$, leads to
\begin{equation}  \label{detstart}
R_{B_{N}^{L}(x),B_{0}^{L}(x)} =
W_{B_{N}^{L}(x),B_{n_S}^{L}(x)}\,\cdot\ldots\cdot \, 
W_{B_{n_2}^{L}(x),B_{n_1}^{L}(x)}
R_{B_{n_1}^{L}(x),B_{0}^{L}(x)} 
\end{equation}
for the given pair $(v,E+\i\eta)$.
For each $1\le n \le N-1$ we ``tile''
-- as indicated in Figure \ref{box2} --
the frame of $B_n^{L}(x)$ with a total number
\begin{equation}\label{tiling}
\tau_n^{L} \le 2d\left( 4nL(N/L)^{1/\nu}+1 \right)^{d-1}
\end{equation}
of boxes $B_0^{\ell}(y)$, whose centres $y$ define the set
$T_n^{L}(x)$, that is,
$\supp \indfkt{B_n^{L}(x)}^{\partial} \subseteq
\overline{\bigcup_{y\in T_n^{L}(x)} \Lambda_{\ell/(4N)}(y) }$.
Hence,
$ \indfkt{B_n^{L}(x)}^{\partial}\le \sum_{y\in T_n^{L}(x)}
\indfkt{B_N^{\ell}(y)}\indfkt{B_0^{\ell}(y)} $
and one deduces for $n<n'\le N$ with the help of Remarks \ref{RWRems}
\begin{align}  \label{detreg}
\bignorm{ W_{B_{n'}^{L}(x),B_{n}^{L}(x)} } & \le
\sum_{y\in T_n^{L}(x)}
\bignorm{R_{B_{n'}^{L}(x),B_{N}^{\ell}(y)} \indfkt{B_0^{\ell}(y)} }
\notag \\
& \le \sum_{y\in T_n^{L}(x)}
\bignorm{W_{B_{n'}^{L}(x),B_{N}^{\ell}(y)}}
\bignorm{R_{B_{N}^{\ell}(y) , B_0^{\ell}(y)} } \; .
\end{align}
This motivates the following two definitions.
\begin{definition} \label{nonrdef}
Let $w\ge 0$.
The potential $ v $ is said to be $ (w,E,L,x) $-\emph{non-resonant}, if
\begin{subequations}
\begin{equation}
\max_{1\le n \le N-1}\; \sup_{\eta>0}
\bignorm{R_{B_{n}^{L}(x),B_{0}^{L}(x)}(v,E+\i\eta)} \le L^{w}
\end{equation}
and
\begin{equation}
\max_{1\le n < n' \le N} \max_{y\in T_{n}^{L}(x)} \; \sup_{\eta>0}
\bignorm{W_{B_{n'}^{L}(x),B_{N}^{\ell}(y)}(v,E+\i\eta)} \le L^{w} \; ,
\end{equation}
\end{subequations}
where $ \ell := (L/N)^{1/\nu} $.
\end{definition}
\begin{definition}
The potential $ v $ is said to be $ (r,E,L,S,x) $-\emph{frame-regular},
if there exist $ S $ natural numbers $ 1 \le n_{1} < \ldots < n_{S}
\le N -1 $ such that $ v $ is $ (r, E, \ell , y) $-regular for all
$ y\in T_{n_{s}}^{L}(x) $ and all $ 1\le s\le S $,
where $ \ell := (L/N)^{1/\nu} $.
\end{definition}
Up to this point we were only concerned with a single potential $v$. 
In order to allow for another potential $v'$, which is supposed to have 
the same properties as $v$,  we use the first resolvent equation
\begin{equation} \label{potchange}
R_{B_{N}^{L}(x),B_{0}^{L}(x)}(v',z)=
R_{B_{N}^{L}(x),B_{0}^{L}(x)}(v,z) +
D_x^{L}(v',v,z) \; ,
\end{equation}
where $ \Im z \not= 0 $ and $D_x^{L}$ is defined in
\begin{definition} \label{appdef}
The potential $ v $ is said to be an $ (r,E,L,x) $-\emph{approximation} of the
potential $ v' $ if
\begin{equation}
\sup_{\eta>0}\bignorm{D_x^{L}(v',v,E+\i\eta)} \le \;\frac{L^{-r}}{2} \, .
\end{equation}
where
\begin{equation}
D_x^{L}(v',v,z):=
R_{B_{N}^{L}(x),(\Lambda_{L}(x),0)}(v,z)\; \big( v-v' \big)
\left( H_{\Lambda_{L}(x)}(v') -z \right)^{-1}
\indfkt{B_{0}^{L}(x)} \; .
\end{equation}
\end{definition}

Combining (\ref{potchange}) with (\ref{detstart}) -- (\ref{detreg}) we
summarize the deterministic part of the recursion clause of the multi-scale
analysis in

\begin{lemma}[Deterministic Part]\label{detPart}
Let $ (S- \nu) r > \nu w(S+1) + \linebreak[4] (d-1)(\nu -1)S $ and assume $ L $
to be sufficiently large. If $ v $ is $ (r,E,L,S,x) $-frame-regular,
$ (w,E,L,x) $-non-resonant and an $ (r,E,L,x) $-approximation of $ v' $,
then $ v' $ is $ (r,E,L,x) $-regular.
\end{lemma}

Next we come to the probabilistic part of the recursion clause of the 
multi-scale analysis. For this purpose let $\tilde{V}$ be a random
potential on $\rz^{d}$ in the sense of Definition~\ref{1.2.4}.
In order to allow for spatially correlated random potentials
it is necessary to control the probability for the joint occurrence of
spatially separated events. More precisely, we have to ensure that
this probability decreases sufficiently fast with increasing
separation.

\begin{definition} \label{KLindep}
Let $K\ge 2$ integer and $\varrho,\vartheta >0$. A random potential
$ \tilde{V} $ is said to be 
$(K,L,\varrho ,\vartheta )$-\emph{independent}, if  
\begin{equation} \label{KLindepglg}
\mathbb{P} \left( \bigcap_{k=1}^{K} A_{k} \right) \le
(L/N)^{-K\vartheta \varrho /\nu }\; 
\end{equation}
for all local events $ A_{k} \in \mathcal{A}_{\tilde{V}}(\Lambda_k)$, 
$ k=1, \ldots , K $, which satisfy
$ \mathbb{P}(A_{k}) \le (L/N)^{-\varrho/\nu } $, 
and all Borel sets $\Lambda_k \subset\rz^d$ subject to 
$ |\Lambda_{k}| \le  (L/N)^{d/\nu}  $
and 
\begin{equation}
\dist_{\infty}(\Lambda_{k}, \Lambda_{k'})
\ge L/(4N) \,\; , \; \text{ for } k\not= k'
\; .
\end{equation}
The sub-sigma-algebra $\mathcal{A}_{\tilde{V}}(\Lambda_k)$ was defined
in Definition \ref{strongmix}.
\end{definition}

\begin{lemma}[Probabilistic Part]\label{probPart}
Let $2\le S\le N-2$ and let $ \big( (N-S)\vartheta -\nu \big) \varrho
> (\nu -1)(d-1)(N-S) $. 
Assume $L$ to be sufficiently large such that, among others,
$\ell:=(L/N)^{1/\nu } \ge 4^{1/(\nu -1)}$.
If $ \tilde{V} $ is
$ (N-S,L ,\varrho ,\vartheta ) $-independent and 
\begin{equation}\label{indVor}
\bbP\big\{ \omega : \tilde{V}^{(\omega)} \text{ is } (r,E,\ell,x)
\text{-regular }\big\}
\ge 1-\ell^{ -\varrho }
\end{equation}
for all $x \in \rz^{d}$, then
\begin{equation}
\bbP\big\{ \omega : \tilde{V}^{(\omega)} \text{ is } (r,E,L,S,x)
\text{-frame-regular }\big\}
\ge 1-L^{ -\varrho }/2
\end{equation}
for all $x \in \rz^{d}$.
\end{lemma}

\begin{proof}
We will estimate from above the probability for the event that
$\tilde{V}$ is \emph{not} $(r,E,L,S,x)$-frame-regular.
Introducing the event
\begin{equation}
A(y):=\{ \omega : \tilde{V}^{(\omega)} \text{ is not } (r,E,\ell,y)
\text{-regular }  \} \;\in\; \mathcal{A}_{\tilde{V}}(\Lambda_l (y))\,,
\end{equation}
one gets from elementary set-theoretic algebra
\begin{eqnarray}
\lefteqn{
\bbP\big\{ \omega : \tilde{V}^{(\omega)} \text{ is not } (r,E,L,S,x)
\text{-frame-regular }\big\}}\nonumber\\
&\le& \sum_{1\le n_{1} < \ldots < n_{N-S} \le N-1}
\bbP \left( \bigcap_{k=1}^{N-S} 
\left( \bigcup_{{\;y\smash{_{k}}\in \; T\smash{_{n_{k}}^{L }}}} 
A(y_{k}) \right) \right)
\nonumber\\ 
&\le&
\sum_{1\le n_{1} < \ldots < n_{N-S} \le N-1} \;
\sum_{y_{1}\in \; T_{n_{1}}^{L }} \ldots
\sum_{y_{N-S}\in \;T_{n_{N-S}}^{L } } 
\bbP \left( \bigcap_{k=1}^{N-S} A(y_{k}) \right) \,. \label{monster}
\end{eqnarray}
By assumption one has $\bbP \bigl(A(y_k)\bigr) \le \ell^{-\varrho}$,
$|\Lambda_\ell (y_k)| = \ell^d$ and 
$\dist_\infty \Big( \Lambda_\ell (y_k),\Lambda_\ell (y_{k'}) \Big)
\ge L/(2N) - \ell \ge L/(4N)$ for $k \not= k'$. Thus, the
$ (N-S,L ,\varrho ,\vartheta ) $-independence of $\tilde{V}$ implies
\begin{equation}
\bbP \left( \bigcap_{k=1}^{N-S} A(y_{k}) \right)
\le (L/N)^{-(N-S)\varrho \vartheta/\nu} \; .
\end{equation}
With the help of this inequality and (\ref{tiling}) one gets an 
$(n_1,\ldots,n_{N-S})$-independent upper bound
for the sums over $y_1,\ldots,y_{N_S}$ in the last line of 
(\ref{monster}). The remaining multiple sum gives a binomial coefficient. 
Hence
\begin{eqnarray}
\lefteqn{
\bbP\big\{ \omega : \tilde{V}^{(\omega)} \text{ is not } (r,E,L,S,x)
\text{-frame-regular }\big\}}\nonumber\\
&\le& \!\!\!\binom{N-1}{N-S} (2d)^{N-S}
\left[ 4(N-1)\, L \left(\frac{N}{L}\right)^{1/\nu} \!+1 \right]^{(d-1)(N-S)}
\left(\frac{N}{L}\right)^{(N-S) \varrho \vartheta/\nu} \nonumber\\
&\le& L^{-\varrho}/2\,,
\end{eqnarray}
if $L$ is sufficiently large.
\end{proof}

The main technical result of this paper, Theorem \ref{multThm} below,
is a multi-scale analysis for truly continuum correlated random
potentials. More precisely, it involves a sequence $ \{V_{k} 
\}_{k\in\bbN_{\,0}} $ of 
random potentials which converges suitably to a random potential $V$.

\begin{figure}[t]
\begin{center}
\parbox{10cm}{\input{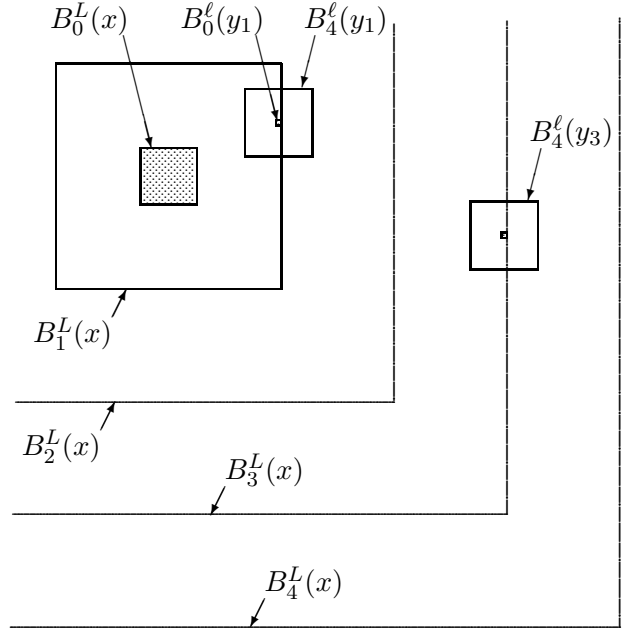}}
\vspace{3ex}
\caption{\label{box3} 
Sketch of the geometric situation considered in Theorem \ref{multThm} for
$d=2$, $N=4$ and $S=2$. The points $y_1$ and $y_3$ are elements of
the sets $T_1^{L}(x)$, respectively $T_3^{L}(x)$.}
\end{center}
\end{figure}

\begin{theorem}\label{multThm}
  Let $V$ and $V_{k}$, $k\in\bbN_{\,0}$, be random potentials on $\rz^{d}$ in
  the sense of Definition~\ref{1.2.4}. Assume that the underlying probability
  space is the same for all these random potentials and that the 
  assumptions of Proposition~\ref{Hdef} hold. Let $ N \ge 4 $ and
  $ 2 \le S \le N-2 $ be natural numbers, let
  $  \vartheta  ,\nu ,\varrho , r, w, L_{0} $ be positive constants
  such that $ \nu >1 $, $ L_{0}^{\nu  -1} \ge 4 $ and
  let $ \{ I_{k} \}_{k \in \bbN_{\,0}} $ be a sequence of
  Lebesgue measurable sets such that
  $ I_{k} \subseteq I_{0} \subseteq \rz $ for all $ k \in \bbN_{\,0} $.
  Define a sequence of lengths  $\{L_k\}_{k\in\bbN_{\,0}}$ through
  $L_{k} := N^{ \frac{\nu ^{k}-1}{\nu -1} } L_{0}^{\nu ^{k}}$, that is,
  \begin{equation} \label{lenseq}
    L_{k+1} = N L_k^{\nu}
  \end{equation}
  and assume that the following five conditions hold for all
  $ k \in \bbN_{\,0} $:
  \begin{indentnummer}
  \item \label{firstass}
    $ (S -\nu ) r > (\nu -1)(d-1)S + \nu w (S+1)  $.
  \item \label{vorindepass}
    $ \big( (N-S) \vartheta -\nu \big) \varrho > (\nu -1)(d-1)(N-S) $.
  \item \label{indepass} 
    $V_{k}$ is $(N-S,L_{k} ,\varrho ,\vartheta )$-independent.
  \item \label{nonruappass}
    For all $ E\in I_{0} $, for all $x\in \rz^{d}$ 
    and for both $ U^{(\omega)}=V^{(\omega)}_{k+1} $ and 
    $ U^{(\omega)}=V^{(\omega)} $ one has
    \begin{align*}
      \bbP\big\{ \omega : \;\, &
      V^{(\omega )}_{k} \text{ is } (w,E,L_{k},x)\text{-non-resonant and} \\
      & \text{an }(r,E,L_{k},x)\text{-approximation of } U^{(\omega )}\; \big\}
      \ge 1-L_{k}^{-\varrho}/2\,.
    \end{align*}
  \item \label{initass}
    For all $ E\in I_{k}$ and for all $x\in \rz^{d}$ one has
    \begin{equation*}
      \bbP\big\{ \omega : \;\, V^{(\omega )}_{k} \text{ is } 
      (r,E,L_{k},x)\text{-regular }\big\} \ge 1-L_{k}^{-\varrho}\,.
    \end{equation*}  
  \end{indentnummer}
  Then there is a $ k_{0} \in \bbN $ such that for all $ k > k_{0} $,
  for all $ E \in I_{k_{0}} $ and for all $ x \in \rz^{d}$ one has
  \begin{equation}\label{multEqn}
    \bbP\big\{ \omega : V^{(\omega)} \text{ is } (r,E,L_{k},x)
    \text{-regular }\big\}
    \ge 1-L_{k}^{ -\varrho }\,.
  \end{equation}
\end{theorem}

\begin{remark}
The subsequent proof shows that \eqref{multEqn} also holds with
$V^{(\omega)}$ replaced by $V_k^{(\omega)}$. In either case 
one should notice that for all $k > k_0$ Eq.~\eqref{multEqn} holds for
all energies $E$ in the fixed set $I_{k_0}$, whereas 
the so-called Initial-Estimate Assumption~(v)
is only required to hold for energies in sets $I_k$ which may become
gradually smaller with increasing $k$. 
\end{remark}

\begin{proof}[Proof of Theorem \ref{multThm}]
Theorem \ref{multThm} is proven by induction on $k$.  
Pick $k_0 \in\bbN$ such that $L:=L_{k_0 +1}$ is large enough as required for 
the applicability of Lemmas~\ref{detPart} and \ref{probPart}. Let 
$U^{(\omega)}$ stand for either $V^{(\omega)}$ or $V_{k_0 +1}^{(\omega)}$.
Due to Assumption~(i) we can apply Lemma~\ref{detPart} with 
$v=V_{k_0}^{(\omega)}$ and $v'=U^{(\omega)}$. For given $E\in\rz$ and given
$x\in\rz^d$ this yields 
\begin{eqnarray}
\lefteqn{\bbP\big\{ \omega : U^{(\omega )} \text{ is } 
(r,E,L_{k_0 +1},x)\text{-regular }\big\} }\nonumber\\
&& \ge -1 + \bbP\big\{ \omega : 
V^{(\omega )}_{k_0} \text{ is } (r,E,L_{k_0 +1},S,x)\text{-frame-regular }
\big\} \nonumber\\
&& \phantom{\ge} + \,\bbP\big\{ \omega : 
V^{(\omega )}_{k_0} \text{ is } (w,E,L_{k_0 +1},x)\text{-non-resonant and} 
\nonumber\\
&& \hspace*{2.3cm} \text{an }(r,E,L_{k_0 +1},x) \text{-approximation of } 
U^{(\omega )}\; \big\}\,.
\end{eqnarray}  
Thanks to Assumptions~(ii), (iii) and (v) we thus conclude from 
Lemma~\ref{probPart} and (iv) that for all $E\in I_{k_0}$ and all
$x\in\rz^d$ 
\begin{equation} \label{msfinal}
\bbP\big\{ \omega : U^{(\omega )} \text{ is } 
(r,E,L_{k_0 +1},x)\text{-regular }\big\} \ge 1-L_{k_0 +1}^{-\varrho}\,.
\end{equation} 
Choosing $U^{(\omega)}=V^{(\omega)}$ in \eqref{msfinal}, we obtain 
\eqref{multEqn} for $k=k_{0}+1$. Choosing $U^{(\omega)}=V^{(\omega)}_{k_0+1}$
in \eqref{msfinal}, we get for all $E\in I_{k_0}$ the substitute for (v) 
on the length scale $L_{k_0 +1}$ which allows one to repeat the above procedure
inductively. 
\end{proof}
%
%
%
\subsection{Macroscopic Limit and Absence of the Absolutely Continuous
  Spectrum}
\label{infvol}
%
%
Starting from the multi-scale estimates (\ref{multEqn}), one can
try to calculate a variety of quantities, describing dynamical localization 
properties, e.g.\ conductivities in the sense of Kubo or diffusion exponents
\cite{FrSp83,MaHo84,GeDBi98,AiGr98,BaFi99,DaSt99}. Here we shall concentrate 
on spectral properties. The
assertion of the following theorem for the infinite-volume random
Schr\"odinger operator $H(V)$ allows one to exclude the absolutely
continuous spectrum for the energies under consideration and also serves as a
starting point for the exclusion of the singular continuous spectrum
with the help of the methods of Theorem~\ref{pp}.

\begin{theorem}\label{macLimitThm}
Assume the situation of Proposition~\ref{Hdef}. 
Fix $N\in\bbN$, a Lebesgue measurable set $I\subseteq\rz$, $L_0 >1$,
$\nu>1$ and set  
$ L_k:=N^{\frac{\nu^{k}-1}{\nu-1}} L_0^{\nu^{k}} $ for $k \in \bbN $.
Assume further that
\begin{equation}\label{macLimitAss}
\bbP\bigl\{ \omega: V^{(\omega)} \text{ is } (r,E,L_k,0)\text{-regular }
\bigr\} \ge 1-L_k^{-\varrho}
\end{equation}
and
\begin{equation} \label{macLimitAss2} 
\bbP\left\{ \omega: {\rm dist}\Big( {\rm spec} \big(H_{\Lambda_{L_k} } 
(V^{(\omega)})\big), E \Big) \le L_k^{-m} \right\}
\le L_k^{-\mu} \tilde{W} 
\end{equation}
holds for all $k\in \bbN$ and all $E\in I$ with suitable constants 
$r, \varrho,m,\mu, \tilde{W} > 0$, obeying $r>4m\nu $.
Then, for all functions $\varphi \in \RL^{2}(\rz^{d})$ obeying
$|\varphi(x)|\le \varphi_0 (1+|x|)^{-\beta}$ for Lebesgue-almost all
$x\in\rz^{d}$ with some constants $\varphi_0<\infty $ and
$\beta>2 m \nu^{2}$, the inequality
\begin{equation} \label{macLimitEqn}
\sup_{\eta > 0 } \bigl\| \bigl( H(V^{(\omega )}) -E -\i \eta 
\bigr)^{-1} \varphi \bigr\| < \infty 
\end{equation}
holds
for \textit{Lebesgue} $\otimes $ $ \bbP$-almost all pairs
$ (E,\omega) \in I \times \Omega $.
\end{theorem}

\begin{proof}
We show that the theorem is a special case of Corollary~2.2 
in \cite{CoHi94}. 
The inequalities $r>4m\nu $ and $\beta>2m\nu^{2} $ imply the existence of 
$r'<r$ with $ 4\,m\nu < r' < 2\,\beta/\nu $. Then estimate (\ref{macLimitAss})
still holds when $r$ is replaced by $r'$.
Now we choose the quantities
$l_k, \varepsilon_k, f $ of \cite{CoHi94} according to
$l_k=L_k $, $\varepsilon_k=L_{k-1}^{-r'/2} $, $f(t)=N^{-m}t^{2m\nu/r'} $.
The sequences $l_k $, resp.\ $ \varepsilon_k$, are monotone increasing, resp.\
decreasing with $\lim_{k\to\infty}\varepsilon_k=0 $.
Since $r'>4\,m\nu$ one has $1/f \in \RL_{\textrm{loc}}^{2}(\rz )$.
The sequences $L_k^{-\mu}$, $L_k^{-\varrho}$ and
$\varepsilon_{k+1}^{-1}l_{k-1}^{-\beta}$ are summable because
$\mu, \varrho >0 $ and $\beta>r'\nu/2$.  Hence, all the
assumptions of Corollary~2.2 in \cite{CoHi94} are satisfied such that
\eqref{macLimitEqn} follows $\bbP$-almost surely for all $E\in I$. Since the 
resolvent of $H(V^{(\omega)})$ is jointly measurable in $E$ and
$\omega$, the subset of pairs $(E,\omega)\in I\times\Omega$ for which
\eqref{macLimitEqn} holds is measurable with respect to the 
product-sigma-algebra and has full $\mathit{Lebesgue}\otimes\bbP$-measure
by Fubini's theorem. 
\end{proof}

\begin{corollary}
Under the assumptions of Theorem~\ref{macLimitThm} 
the random Schr\"odinger operator $H(V)$
has $\bbP$-almost surely no absolutely continuous spectrum in $I$. 
\end{corollary}

\begin{proof}
Let $\{u_n\}_{n\in\bbN}$ be a complete orthonormal sequence of vectors in
$\RL^2(\rz^d)$ and suppose that $|u_n(x)|\le u_{n,0} (1+|x|)^{-\beta}$
for all $x\in\rz^{d}$ with some constants $u_{n,0} < \infty$ and $\beta$ as in
Theorem~\ref{macLimitThm}. 
It suffices to show that the event
\begin{equation}
\bigl\{ \omega :\, \langle u_n, F_{\mathrm{ac}}(V^{(\omega)}, I) u_n\rangle
=0 \text{ for all } n\in\bbN \bigr\} 
\end{equation}   
occurs with probability one. Here, $\langle\cdot,\cdot\rangle$ is
the scalar product \eqref{scalarproduct} on $\RL^{2}(\rz^{d})$, and 
$F_{\mathrm{ac}}(V^{(\omega)},\cdot)$
denotes
the absolutely continuous component arising in the Lebesgue decomposition 
of the projection-valued spectral measure of $H(V^{(\omega)})$. But
this follows from  
\begin{eqnarray}
\lefteqn{\int_{\Omega}\!\d\bbP(\omega)\; 
\langle u_n, F_{\mathrm{ac}}(V^{(\omega)}, I) u_n\rangle }\nonumber\\
&& = \frac{1}{\pi}\; \int_{\Omega}\!\d\bbP(\omega)\; \left(
\int_I\!\d E\; \lim_{\eta\downto 0} \Im\,\bigl\langle u_n, 
\bigl(H(V^{(\omega)}) -E -\i\eta\bigr)^{-1} u_n\bigr\rangle\right)
\nonumber\\
&& = \frac{1}{\pi}\; \int_{I\times \Omega}\!\d E\otimes\d\bbP(\omega)\; 
\lim_{\eta\downto 0}\left( \eta\,\bignorm{\bigl(
H(V^{(\omega)}) -E -\i\eta\bigr)^{-1} u_n}^2\right)\nonumber\\
&& = 0\,,
\end{eqnarray}
where we have used the Theorem of Fatou and de la Vall\'ee Poussin,
see e.g.\ Thm.~A.10 in \cite{PaFi92}, Fubini's Theorem and 
\eqref{macLimitEqn}.
\end{proof}

%

\section{Pure-Point Spectrum}  \label{ppSec}

From the last section we know that the results of the multi-scale analysis 
for energies in $I\subset\bbR$ suffice to exclude the existence of 
the absolutely continuous spectrum of the infinite-volume random
Schr\"odinger operator $H(V)$ in $I$ 
with probability one. In order to exclude the singular continuous
spectrum as well, and hence to show that the  
spectrum of $H(V)$ is $\mathbb{P}$-almost surely only pure point in $I$,
additional assumptions and efforts are needed. This is accomplished 
by Theorem \ref{pp} below, which adapts and elaborates on some results  
of Simon and Wolff \cite{SiWo86}, Howland \cite{How87} and Combes and 
Hislop \cite{CoHi94} in order to be applicable to truly continuum
random potentials admitting a rather general one-parameter decomposition.

\begin{theorem} \label{pp}
  Assume the situation of Proposition~\ref{Hdef}. Furthermore, let $n>
  d/4$ be  a 
  natural number such that 
  \begin{indentnummer}
  \item\label{ppmom} 
    $\mathbb{E} \bigl\{ |V(0)|^{2n} \bigr\} =: c _{n} < \infty$,
  \item\label{ppdecomp} 
    $V$ admits a one-parameter decomposition 
    \begin{equation*} 
      V^{(\omega )} = U^{(\omega )} + \lambda ^{(\omega )}  u
    \end{equation*}
    for $\mathbb{P}$-almost all $\omega$, where $U: \Omega \times\rz^d 
    \rightarrow \rz$ is a random field (which, in general, is
    not homogeneous),  
    $u \in \mathrm{L}^{n}(\rz^d) \cap \mathrm{L}^{\infty }(\rz^d)$ 
    is a strictly positive function and
    $\lambda : \Omega \rightarrow \rz$ is a random variable whose
    conditional probability measure relative to the sub-sigma-algebra
    $\mathcal{A}_{U}\subseteq\mathcal{A}$ generated by
    $\{U(x)\}_{x\in\rz^{d}}$ has a density $\rho :
    \Omega\times\rz\rightarrow\rz^{+}$ with respect to Lebesgue
    measure, which is  measurable with respect to the product
    sigma-algebra $\mathcal{A}_{U}
    \otimes\mathcal{L}$. Here, $\mathcal{L}$ is the sigma-algebra of
    Lebesgue measurable sets in $\rz$,
  \item\label{ppnorm} 
    there exists $I \in \mathcal{L}$ 
    such that for $\mathbb{P}$-almost all $\omega \in \Omega$ one 
    can find $I_{0} \in \mathcal{L}$ with $I_{0}(\omega )\subseteq I$ and 
    $| I\!\setminus \! I_{0}(\omega )| = 0$ such that 
    \begin{equation*} 
      \sup_{\eta > 0 } \bigl\| \bigl( H(V^{(\omega )}) -E -\i
      \eta 
      \bigr)^{-1} u^{1/2} \bigr\| < \infty  \qquad \text{for all\;\;}
      E\in  I_{0}(\omega )\,.
    \end{equation*}
  \end{indentnummer}
  \vspace{-3ex}
  Then, the spectrum of $H(V)$ is $\mathbb{P}$-almost surely only pure
  point in $I$. 
\end{theorem}

The proof of Theorem~\ref{pp} relies on a deterministic result about
the instability of  the singular continuous spectrum under
perturbations. We quote a special case of Thm.~3.2 in \cite{CoHi94} as

\begin{proposition} \label{singcont}
  Let $ h_{0} $ be a self-adjoint operator with domain $\mathcal{D}(h_{0}) 
  \subseteq \mathrm{L}^2(\rz^d) $. For a bounded, non-negative,
  self-adjoint operator
  $ h_{1} $ and $ \xi \in\rz $ define $ h(\xi ) :=
  h_{0} + \xi h_{1} $ on $ \mathcal{D}(h_{0}) $ and assume that
  \begin{indentnummer}
  \item\label{singcontcomp} 
    there is $J \in\mathcal{L}$ such that $ h_{1}^{1/2} (
    h_{0} -E -\i\eta )^{-1} h_{1}^{1/2} $ is compact for all 
    $ \eta > 0 $ and all $ E \in J $,
  \item\label{singcontnorm} 
    there is $J_{0} \in\mathcal{L}$ with $ J_{0}\subseteq J $ and 
    $ | J\!\setminus\! J_{0} | =0 $ such that 
    \begin{equation*} 
      \sup_{\eta > 0 } \bigl\| ( h_{0} -E -\i \eta )^{-1} 
      h_{1}^{1/2} \bigr\| < \infty  \qquad \text{for all\;\;}
      E\in  J_{0}\,,
    \end{equation*}
  \item\label{singcontdense} 
    the subspace $ \{ h_{1}^{1/2}\psi : \psi \in \mathrm{L}^2(\rz^d) \} $
    is dense in $ \mathrm{L}^2(\rz^d) $.
  \end{indentnummer}
  Then, for Lebesgue-almost all  $\xi \in\rz$ the spectrum of 
  $h (\xi )$ is only pure point in $J$ with finitely degenerate
  eigenvalues. 
\end{proposition}

\begin{proof}[Proof of Theorem \ref{pp}]
  Let us define the function $X: \Omega\times\rz\rightarrow\{0,1\}$ by
  setting $X^{(\omega)}(\xi)=0$, if a one-parameter decomposition of
  $V$ exists in the sense of Assumption \itemref{ppdecomp}, the operator 
  $H(U^{(\omega)})+\xi u$ is self-adjoint and its spectrum is pure
  point in $I$. Otherwise we set $X^{(\omega)}(\xi)=1$.
  The joint measurability of the random potential $V$ implies that the
  random potential 
  $\bigl((\omega,\xi), x \bigr) \mapsto U^{(\omega)}(x) + \xi u(x)$
  is jointly measurable
  with respect to the product sigma-algebra $(\mathcal{A}_{U} \otimes
  \mathcal{L}) \otimes \mathcal{B}^{\,d}$, where $\mathcal{B}^{\,d}$ is the
  sigma-algebra of the Borel sets in $\rz^{d}$. Accordingly, 
  \cite{KiMa82} implies the joint measurability of $X$ with respect 
  to the completion $(\mathcal{A}_{U}\otimes\mathcal{L})\,\tilde{~}\,$
  of $\mathcal{A}_{U}\otimes\mathcal{L}$ induced by the product measure
  $\bbP\otimes\mathit{Lebesgue}$.
  We also define the $\Omega$-subsets
  \begin{equation}
    \begin{split}
      \Omega_{1} &:= \bigl\{ \omega\in\Omega : 
      X^{(\omega)}(\lambda^{(\omega)}) = 0 \bigr\}\,,\\
      \Omega_{0} &:= \biggl\{ \omega\in\Omega : \int_{\rz}\!\d\xi\;
      X^{(\omega)}(\lambda^{(\omega)} + \xi) = 0 \biggr\}\,.
    \end{split}
  \end{equation}
  Apart from the $\bbP$-null set allowed for in Assumption
  \ref{ppdecomp}, $\Omega_{1}$ is the set of $\omega$'s for which
  $H(V^{(\omega)})$ enjoys the property of being
  self-adjoint with only 
  pure-point spectrum in $I$. Hence, we know from Proposition
  \ref{Hdefmess} that $\Omega_{1}\in\mathcal{A}$. 
  Analogously, up to the same $\bbP$-null set,
  $\Omega_{0}$ is the set of $\omega$'s such that for
  Lebesgue-almost all $\xi\in\rz$ 
  the operator $H(V^{(\omega)}) + \xi u$ is 
  self-adjoint with only
  pure-point spectrum in $I$. The $\mathcal{A}$-measurability of 
  $\Omega_{0}$ follows from the completeness of $\mathcal{A}$, the joint
  measurability of $X$ and Fubini's theorem.
  
  We split the rest of the proof into two parts. In part~a) we show that
  $\bbP(\Omega_{0}) = 1$ implies $\bbP(\Omega_{1}) = 1$
  and in part~b) that the assumptions of the theorem imply 
  $\bbP(\Omega_{0}) = 1$.

  a)~ Suppose $\bbP(\Omega_{0}) = 1$, then we have 
  \begin{equation}
    0 = \bbE \biggl\{ \int_{\rz}\!\d\xi\;
    X(\lambda + \xi) \biggr\} 
    = \bbE \biggl\{ \int_{\rz}\!\d\xi\; X(\xi) \biggr\}\,,
  \end{equation}
  and Fubini's theorem gives $X^{(\omega)}(\xi) =0$ for almost all pairs
  $(\omega,\xi)\in 
  \Omega\times\rz$ with respect to the completed measure
  $(\bbP\otimes\mathit{Lebesgue})\,\tilde{~}\,$. It follows the existence
  of $Y: \Omega\times\rz\rightarrow \{0,1\}$ such that $X\le Y$, $Y$ is 
  $\mathcal{A}_{U}\otimes\mathcal{L}$-measurable and
  $X^{(\omega)}(\xi) = Y^{(\omega)}(\xi)$ for 
  $\bbP\otimes\mathit{Lebesgue}$ almost all $(\omega,\xi)\in
  \Omega\times\rz$. Indeed, the level set $\{ (\omega,\xi)\in
  \Omega\times \rz : X^{(\omega)}(\xi) =0\} \in
  (\mathcal{A}_{U}\otimes\mathcal{L})\,\tilde{~}\,$ differs at most by
  a $(\bbP\otimes\mathit{Lebesgue})\,\tilde{~}\,$-null set from a
  product measurable set
  $\Xi\in\mathcal{A}_{U}\otimes\mathcal{L}$. Now define
  $Y^{(\omega)}(\xi) =0$ for all $(\omega,\xi) \in \Xi$ and
  $Y^{(\omega)}(\xi) =1$ elsewhere. Thus, we conclude from Fubini's
  theorem that  
  \begin{equation}
    \label{disintegration}
    0 = \bbE \biggl\{ \int_{\rz}\!\d\xi\; \rho(\xi)\,Y(\xi) \biggr\}
    = \bbE\Bigl\{ \bbE\bigl\{ Y(\lambda)\,|\,\mathcal{A}_{U} \bigr\}\Bigr\} 
    = \bbE\{ Y(\lambda) \}\,.
  \end{equation}
  The second equality in \eqref{disintegration} follows from a slight
  generalization of the ``Disintegration Theorem~5.4'' in \cite{Kal97}
  and uses the fact that the
  $\mathcal{A}_{U}\otimes\mathcal{L}$-measurable Lebesgue density 
  $\rho$ provides a regular version of the
  conditional probability measure of $\lambda$ for given
  $\mathcal{A}_{U}$.  
  Hence, $Y^{(\omega)}(\lambda^{(\omega)})=0$ for $\bbP$-almost all 
  $\omega\in\Omega$. Since $0 \le X \le Y$ this implies
  $\bbP(\Omega_{1})=1$.

  b)~ The aim is to derive $\bbP(\Omega_{0})=1$ from the conclusion of
  Proposition~\ref{singcont}. To this end we have to ensure that the
  assumptions of Proposition~\ref{singcont} are satisfied for
  $\bbP$-almost all $\omega\in\Omega$ when setting $h_{1}= u$, $J=I$
  and $h_{0}= H(V^{(\omega )})$. 
  
  Recall that by assumption, $u$ is a strictly positive, bounded
  function and that by Proposition \ref{Hdefess} 
  $H(V)$ is $\bbP$-almost surely essentially self-adjoint on
  $\mathcal{C}_{0}^{\infty}(\rz^{d})$. 
  Obviously, Assumption \ref{ppnorm} is identical to Assumption
  \ref{singcontnorm}, if we also set $J_{0}= I_{0}(\omega )$.

  As to Assumption \ref{singcontdense}, suppose that $\varphi$ is in the 
  orthogonal complement $\mathcal{H}^{\bot}$ of the norm-closed subspace
  $\mathcal{H}:= \{ u^{1/2}\psi : \psi \in
  \mathrm{L}^2(\rz^d)\}^{\mathrm{cl}}$ of 
  $\mathrm{L}^2(\rz^d)$. Since $u^{1/2}\varphi \in \mathcal{H}$ it
  follows that 
  \begin{equation} 
    0= \langle \varphi , u^{1/2}\varphi\rangle =
    \int_{\rz^d}\!\d^d x\; |\varphi (x)|^2 \bigl( u(x)\bigr)^{1/2}\,.
  \end{equation}
  But $u(x)> 0$ for Lebesgue-almost all $x\in\rz^d$ implies 
  $\varphi (x) =0$ for Lebesgue-almost all $x\in\rz^d$  so that 
  $\mathcal{H}^\bot =\{0\}$ and $\mathcal{H}=\mathrm{L}^2(\rz^d)$, proving 
  \ref{singcontdense}.

  It remains to verify Assumption \ref{singcontcomp}, that is, the
  $\bbP$-almost sure compactness of 
  $u^{1/2}\bigl( H(V) -E -\i\eta \bigr)^{-1} u^{1/2}$
  for all $\eta > 0$ and all $E\in I$. 
  Because of the boundedness of $u$ it is sufficient to show that  
  the operator 
  \begin{equation} \label{relcomp}
    u^{1/2}\bigl( H(V^{(\omega )}) -z \bigr)^{-1} 
  \end{equation}
  is compact for $z$ equal to some fixed $z_{0}\in
  \cz\!\setminus\!\rz$ for $\mathbb{P}$-almost all $\omega
  \in\Omega$. Note that  
  compactness of (\ref{relcomp}) for some $z_{0}$
  implies compactness of (\ref{relcomp}) for all $z\in \cz\!\setminus\!\rz$.
  Hence the set of $\omega$'s for which (\ref{relcomp}) is 
  compact does not depend on $z\in \cz\!\setminus\!\rz$. By the second 
  resolvent equation
  \begin{multline}  \label{sre}
    u^{1/2}\bigl( H(V^{(\omega )}) -z_{0} \bigr)^{-1} =
    u^{1/2}\bigl( H(0) -z_{0} \bigr)^{-1}\\ - 
    u^{1/2}\bigl( H(0) -z_{0} \bigr)^{-1} V^{(\omega )}
    \bigl( H(V^{(\omega )}) -z_{0} \bigr)^{-1}\,,
  \end{multline}
  it suffices in turn to prove compactness of 
  $u^{1/2}\bigl( H(0) -z_{0} \bigr)^{-1}$ and of
  $u^{1/2}\bigl( H(0) -z_{0} \bigr)^{-1} V^{(\omega )}$ for 
  $\mathbb{P}$-almost all $\omega \in\Omega$. We note that the
  validity of the resolvent equation \eqref{sre} itself is ensured by
  this compactness and the fact that \eqref{sre} obviously holds on
  the subspace $\bigl\{ \psi\in\RL^{2}(\rz^{d}) : \psi = \bigl(
  H(V^{(\omega)}) -z_{0}\bigr)\phi,
  \phi\in\mathcal{C}_{0}^{\infty}(\rz^{d})\bigr\}$ which, according to
  the Cor.\ on p.~257 in \cite{ReSi80} is dense in $\RL^{2}(\rz^{d})$,
  because $H(V^{(\omega)})$ is essentially self-adjoint on
  $\mathcal{C}_{0}^{\infty}(\rz^{d})$. 

  To show this compactness put $z_{0}= \tilde{E} +\i\eta$ with
  $\tilde{E} < 0$, 
  $\eta \neq 0$, and observe the diamagnetic inequality (see e.g.\
  \cite{CyFr87}) for the
  integral kernels of the resolvents of $H(0)$ and $ - \Delta /2$  
  \begin{equation} 
    \bigl| \bigl(H(0) -z_{0}\bigr)^{-1}(x,y)\bigr| \le
    \bigl(- \Delta /2 -\tilde{E}\bigr)^{-1}(x,y) =: G_{1}(x-y)\,,
  \end{equation}
  which are jointly continuous on $\{(x,y)\in\rz^{d}\times\rz^{d}:
  x\neq y\}$.
  Let $n> d/4$ be the natural number defined in Assumption \ref{ppmom}.
  Then, we have for both $j=0$ and $j=1$ the inequality
  \begin{align} \label{trace1}
    \mathbb{E}&\Bigl\{ \Tr \Bigl\{\bigl[\bigl( u^{1/2}(H(0) -z_{0})^{-1}
    V^{j}\bigr) \bigl( u^{1/2}(H(0) -z_{0})^{-1} V^{j}\bigr)^\dagger
    \bigr]^n\Bigr\}\Bigr\} \notag\\ 
    & \le \mathbb{E} \Biggl\{ \intr[d]{x_{1}}\!\!\intr[d]{y_{1}}
    \ldots\intr[d]{x_{n}} 
    \!\!\intr[d]{y_{n}} \Biggl( \prod_{\nu=1}^n u(x_{\nu})\,
    |V(y_{\nu})|^{2j} \Biggr)\notag\\
    & \qquad\qquad\times G_{1}(x_{1}-y_{1})G_{1}(y_{1}-x_{2})
    \cdot\ldots\cdot G_{1}(x_{n}-y_{n}) G_{1}(y_{n}-x_{1})
    \Biggr\}\notag\displaybreak[0]\\ 
    & \le c_{n}^{j} \!\intr[d]{r_{2}}\!\ldots\!\intr[d]{r_{n}}
    G_{2}(-r_{2}) G_{2}(r_{2}-r_{3})\cdot\!\ldots\!\cdot
    G_{2}(r_{n-1}-r_{n}) G_{2}(r_{n})\notag\\
    & \qquad\qquad \times \intr[d]{r_{1}} u(r_{1})
    u(r_{1}+r_{2}) \cdot\ldots\cdot u(r_{1} + r_{n}) \notag\\
    & \le c_{n}^{j} \, \norm[n]{u}^{n} \, G_{2n}(0)\,.
  \end{align}
  To obtain the second inequality in \eqref{trace1} we have used 
  the iterated H\"older inequality in order to employ Assumption
  \ref{ppmom}, the convolution property 
  \begin{equation}
    \label{convprop}
    G_{k+1}(x-y) = \int_{\rz^{d}}\d^{d}w\; G_{k}(x-w)\, G_{1}(w-y)
  \end{equation}
  for the $k$-times iterated integral kernel $G_{k}(x-y):=\bigl( -
  \Delta /2 - 
  \tilde{E}\bigr)^{-k}(x,y)$ and the change-of-variables 
  $r_{1}:=x_{1}$, $r_{l}:= x_{l} - x_{1}$ for $l=2,\ldots,n$.
  The last inequality in \eqref{trace1} follows from the iterated H\"older
  inequality and \eqref{convprop}. Since $u\in\RL^{n}(\rz^{d})$ and
  $G_{2n}(0)=(2\pi)^{-d}\int_{\rz^{d}} \d^{d}p\;
    \bigl((p^{2}/2)-\tilde{E}\bigr)^{-2n}$, 
  the upper bound \eqref{trace1} is finite because of $n >
  d/4$. Hence, the operator $u^{1/2}(H(0) -z_{0})^{-1} V^{j}$ is
  $\bbP$-almost surely compact for both $j=0$ and $j=1$ by Prop.~6 on
  p.~42 in \cite{ReSi75}.
\end{proof}


\section{Proof of the Main Theorem \ref{maintheo}}\label{applSec}

In this section we prove Theorem~\ref{maintheo} by showing that the
assumptions of the more general theorems of Sections~\ref{multiscale}
and \ref{ppSec} are fulfilled for the Gaussian random potentials 
under consideration. The main result of this section is the following
Theorem~\ref{msaThm}, which establishes multi-scale estimates in the
low-energy and weak-disorder regime. These estimates for the
finite-volume resolvents will then be transported to the
infinite-volume resolvent in Corollary~\ref{resboundedThm} by means of
Theorem~\ref{macLimitThm}. Given Corollary~\ref{resboundedThm}, the
proof of Theorem~\ref{maintheo} is finally completed by an
application of Theorem~\ref{pp}.

\begin{theorem}\label{msaThm}
  Let $V$ be a Gaussian random potential on $\rz^{d}$ with covariance
  function $x\mapsto \sigma^2 C(x)$, where $\sigma>0$ and $C$ has either
  property \ass{R} or the three properties \ass{P}, \ass{H}
  and \ass{M}.  
  Let $H(V)$ be the associated random Schr\"odinger operator as in
  Proposition~\ref{Hdef}.
  Then there are a natural number $N\ge4$ and  positive reals
  $\nu, \varrho, r$ obeying

  \begin{equation}
    \label{msaNu}
    1<\nu<1+\frac{1}{8d}
  \end{equation}
  and
  \begin{equation}\label{rNuD}
    4 \nu d < r
  \end{equation}
  such that  the following two statements hold.
  \begin{indentnummer}
  \item \label{lowpart}
    For every $\sigma > 0 $ there is $1< L_0 <\infty$ and 
    $-\infty < E_0 < 0$
    such that
    \begin{equation}\label{msaThmequ}
      \bbP \bigl\{ \omega : V^{(\omega)} \text{ is } (r,E,L_k,x)\text{-regular}
      \bigr\} \ge 1-L_k^{-\varrho } 
    \end{equation}
    for all $x\in\rz^{d}$ and all $E \in \; ]-\infty,E_0]$, where
    $ L_k:=N^{\frac{\nu^{k}-1}{\nu-1}} L_0^{\nu^{k}} $, $ k\in \bbN $.
  \item  \label{weakpart}
    For every $E_0 < 0 $ there is $1< L_0 <\infty$ and $\sigma_0>0$
    such that for all $\sigma\in\,]0,\sigma_0] $ the
    inequality (\ref{msaThmequ}) holds for all $x\in\rz^{d}$,
    all $ E \in \; ]-\infty,E_0] $ and all $k\in\nz$.
  \end{indentnummer}
\end{theorem}

For the time being let us assume that Theorem~\ref{msaThm} is valid 
and proceed with the proof of Theorem~\ref{maintheo}.

\begin{corollary} \label{resboundedThm}
  Assume the situation of Theorem \ref{msaThm}.  
  Then, for all functions $\varphi\in\RL^{2}(\rz^{d}) $ with
  $|\varphi(x)|\le\varphi_0 (1+|x|)^{-\beta}$ 
  for some constants $\varphi_0<\infty $ and  $\beta>2d +3/4$
  the inequality
  \begin{equation}
    \sup_{\eta > 0 } \bigl\| \bigl( H(V^{(\omega )}) -E -\i \eta 
    \bigr)^{-1} \varphi \bigr\| < \infty 
  \end{equation}
  holds  for \textit{Lebesgue} $\otimes $ $ \bbP$-almost all pairs
  $ (E,\omega) \in \; ]-\infty,E_0] \times \Omega $.
\end{corollary}

\begin{proof}
  We check that the assumptions of Theorem~\ref{macLimitThm} are
  fulfilled for $I:=]-\infty, E_{0}]$. Obviously, \eqref{msaThmequ}
  provides \eqref{macLimitAss}. Since property \ass{R} of a Gaussian
  random potential implies  property \ass{P}, the requirement
  \eqref{macLimitAss2} follows from the Wegner estimate in the weakened
  form \eqref{wegnerloc}, viz.\  
  \begin{equation}
    \bbP\left\{ \omega: {\rm dist}\Big( {\rm spec} \big(H_{\Lambda_{L_k} } 
      (V^{(\omega)})\big), E \Big) \le L_k^{-m} \right\}
    \le 2 L_k^{d-m} W(E_0+ L_{0}^{-m}) 
  \end{equation}
  for all $k\in \bbN$ and all $E\in \; ]-\infty,E_0]$. Thanks to
  \eqref{rNuD} there exists a real constant $m$ such that $d<m<
  \min\{ r/(4\nu), d+(4\nu)^{-2}\}$. Hence, $\mu:=m-d >0$, $r>4m\nu$
  and, since $2m\nu^{2} < 2d +3/4$ according to \eqref{msaNu}, we obtain
  the lower bound $\beta > 2d +3/4$ for the decay exponent $\beta$.
\end{proof}

\begin{proof}[Proof of Theorem \ref{maintheo}]
  We check the assumptions of Theorem~\ref{pp} with $I:=]-\infty,
  E_{0}]$. Assumption \ref{ppmom} is obviously fulfilled for Gaussian
  random potentials for any $n\in\nz$.

  Concerning Assumption \ref{ppdecomp} we introduce a one-parameter
  decomposition of $V$ by defining the function
  \begin{equation}
    u(x) := \mathcal{N}^{-1/2} \intr[d]{y} \e^{-y^{2}/2}\, C(x-y)
  \end{equation}
  for $x\in\rz^{d}$, $\mathcal{N}:= \intr[d]{x}\e^{-x^{2}/2}
  \intr[d]{y} \e^{-y^{2}/2}\, C(x-y)$ being a normalization constant,
  the centred Gaussian random variable
  \begin{equation}
    \lambda^{(\omega)}:= \mathcal{N}^{-1/2} \intr[d]{y}
    \e^{-y^{2}/2}\, V^{(\omega)}(y)
  \end{equation}
  with variance $\bbE(\lambda^{2}) = 1$ and the non-homogeneous
  Gaussian random field
  \begin{equation}
    U^{(\omega)}(x) := V^{(\omega)}(x) - \lambda^{(\omega)} u(x)\, .
  \end{equation}
  Due to the Gaussian nature of both $\lambda$ and $U$, and due to
  $\bbE \{ \lambda U(x)  \} =0$ for all $x \in \rz^{d}$
  we conclude that 
  the random variable $\lambda$ is stochastically independent of
  $\{ U(x) \}_{x \in \rz^{d}}$. Hence, the conditional probability
  density $\xi\mapsto\rho^{(\omega)}(\xi) =
  (2\pi)^{-1/2}\exp\{-\xi^{2}/2\}$ of $\lambda$, given $\{ U(x) 
  \}_{x \in \rz^{d}}$, is independent of $\omega$ and has clearly
  the desired measurability properties. Note that
  $u\in\RL^{\infty}(\rz^{d})$ is a strictly positive function because
  of Remark \ref{zetaC}, property \ass{P} and $C(0)>0$. Moreover, it
  follows from the same remark and property \ass{D} that 
  \begin{equation}
    \label{udecay}
    |u(x)| \le u_{0} (1+|x|)^{-z}
  \end{equation}
  for all $x\in\rz^{d}$, where $0 < u_{0} < \infty$ and $z>
  4d+3/2$. Hence, $u\in\RL^{n}(\rz^{d})$ for any natural number $n >
  d/4$. 

  Finally \eqref{udecay} ensures that $u^{1/2}$ is an admissible
  choice for $\varphi$ in Corollary~\ref{resboundedThm}, and
  Assumption \ref{ppnorm} is seen to hold by Corollary \ref{resboundedThm} 
  and Fubini's theorem. 
\end{proof}

It remains to prove Theorem~\ref{msaThm} by checking that the
assumptions of Theorem~\ref{multThm} hold. This will be done in
Subsection~\ref{proofmsaThm} below. The first three subsections are
devoted to necessary preparations concerning the control of large
fluctuations, the initial estimates and the control of long-range
correlations, respectively.

\subsection{Controlling Large Fluctuations}\label{fluctSec}

Typical realizations of Gaussian random potentials are unbounded. It is
however possible to bound the probability that 
the absolute value of the realizations in a given
cube exceeds a given value . This L\'evy type of ``maximal
inequality'' is the content of 

\begin{lemma}\label{fluctLemma}
  Consider a Gaussian random potential $V$ on $\rz^{d}$ with 
  property \ass{H}. Then, its realizations $x \mapsto V^{(\cdot)}(x)$
  are $\bbP$-almost surely continuous functions on $\rz^{d}$ and there
  exists a length $1 < \ell_{\mathsf{H}} < \infty$ such that
  \begin{equation}\label{fluctInequ}
    \bbP \left\{ 
      \omega: \, \sup_{x\in\Lambda_{\ell }(0)} |V^{(\omega )}(x)| \ge E
    \right\}
    \le
    2^{2(d+1)} \, \exp\left\{ - \frac{E^2}{200 \, C(0) \ln \ell} \right\}
  \end{equation}
  holds for all $\ell \ge \ell_{\mathsf{H}}$ and all $E \ge 0$.
  The length $\ell_{\mathsf{H}}$ depends only on the 
  value of the H\"older exponent and the size of the neighbourhood
  referred to in property \ass{H}.
\end{lemma}

\begin{proof}
  We deduce the lemma from Thm.~4.1.1 in \cite{Fer75}. 
  When adapted to a homogeneous random field and a cube
  $\Lambda_{\ell}(0)$ with edges of length $\ell >0$, this theorem
  implies that, if  
  \begin{equation}\label{ferniqueVor}
    \mathcal{J}(\ell) := \int^{\infty}_1 \Rd p \,
    \sup_{|x|_\infty\le\ell\, 2^{-p^{2}}}
    \sqrt{C(0)-C(x)} \; < \infty\,,
  \end{equation}
  then the realizations of $V$ are $\bbP$-almost
  surely continuous functions on $\Lambda_{\ell}(0)$, and the inequality
  \begin{equation}
    \label{orifer}
    \bbP \left\{ 
      \omega: \, \sup_{x\in\Lambda_{\ell }(0)} |V^{(\omega )}(x)| \ge E
    \right\}
    \le    
    \frac{5}{2}\; 2^{2d}\int_{\xi_{\ell}(v)}^{\infty}\!\d q\;
    \e^{-q^{2}/2}  
  \end{equation}
  holds for all $E\ge 0$ obeying
  \begin{equation}
    \label{orifercond}
    \xi_{\ell}(E) := E\, \Bigl[ \sqrt{C\smash{(0)}} +
    (2+2^{3/2})\mathcal{J}(\ell)\Bigr]^{-1}  \ge \sqrt{1 + 4d\ln 2}\,.
  \end{equation}
  First, we verify that condition \eqref{ferniqueVor} is fulfilled
  for all $\ell >1$ as a consequence of property \ass{H}. To this end
  we introduce $\theta >0$ to characterize the size of the
  neighbourhood referred to in property \ass{H} by $|x|_\infty \le \theta$. 
  Without restriction we may assume that $\theta \le \min\bigl\{ 1, \bigl(
  2C(0)/b\bigr)^{1/\beta}\bigr\}$. Hence, property \ass{H} implies 
  \begin{equation}\label{implH}
    \sup_{|x|_\infty\le\ell\, 2^{-p^{2}} }
    \big( C(0)-C(x) \big)
    \le
    b\left( \ell \, 2^{-p^{2}}\right)^{\beta } 
  \end{equation}
  for all  $p \ge p_{0} := \sqrt{\bigl(\ln(\ell/\theta)\bigr)/\ln 2}$.
  By extending the lower limit of the integration in (\ref{ferniqueVor})
  to zero, splitting the integral into two parts at $p_{0}$
  and using $|C(x)|\le C(0)$ in the first, respectively 
  (\ref{implH}) in the second part, one arrives at the upper bound
  \begin{equation}\label{ferniqueVor2}
    \mathcal{J}(\ell) \le
    \sqrt{2\, C(0)\, \frac{\ln (\ell /\theta) }{\ln 2}\, } +
    \sqrt{\frac{2 b \ell^{\beta}}{\beta \ln 2}}
    \int_{\sqrt{(\beta/2) \ln (\ell/\theta)}}^{\infty}
    \Rd p \, \e^{-p^{2}}\,.
  \end{equation}
  Hence, $\mathcal{J}(\ell)$ is finite and Thm.~4.1.1 in \cite{Fer75}
  is applicable. In order to reduce the right-hand side of
  \eqref{orifer} to the more explicit (but less sharp) bound claimed
  in \eqref{fluctInequ}, we exploit 
  \begin{equation}
    \label{abramineq}
    \int^{\infty}_s \Rd p \;
    \e^{-{p}^{2}} \le \frac{\sqrt{\pi}}{2}\; \e^{-s^{2}}\,,
  \end{equation}
  for $s \ge 0$, which follows from Formula~7.1.13 in \cite{AbSt72}. 
  Using this and
  $\theta^{\beta} \le 2C(0) /b$, we continue to estimate
  (\ref{ferniqueVor2}) from above by
  \begin{equation}\label{ferniqueVor3}
    \mathcal{J}(\ell) \le
    \left\{ \sqrt{\frac{2}{\ln 2}\,
        \left(1-\frac{\ln\theta }{\ln \ell} \right)}+
      \sqrt{ \frac{\pi}{\beta (\ln 2)(\ln \ell)} }
    \right\} \sqrt{C(0)\ln \ell} \; .
  \end{equation}
  The term in curly brackets
  in (\ref{ferniqueVor3}) approaches $\sqrt{2/\ln 2\,} $ for
  $\ell \to \infty$. Thus one can find a length $1<
  \ell_{\mathsf{H}} < \infty$, depending on $\theta $ and $\beta$,
  such that 
  \begin{equation}
    \label{ferlast}
    \sqrt{C(0)\,}+
    (2+2^{3/2}) \mathcal{J}(\ell) 
    \le 10 \sqrt{C(0) \ln \ell}
  \end{equation}
  holds for all $\ell\ge\ell_{\mathsf{H}}$. 
  Upon inserting this inequality and \eqref{abramineq} into
  \eqref{orifer}, we deduce \eqref{fluctInequ} for all $E$ obeying the
  condition \eqref{orifercond}. For those $E$ not obeying
  \eqref{orifercond}, the inequality \eqref{fluctInequ} is trivially true, 
  because then its right-hand side is bigger than one by \eqref{ferlast}.
\end{proof}

\begin{remark}
  The bound \eqref{fluctInequ} is very rough as can be inferred from
  the exact asymptotic behaviour 
  \begin{equation} 
    \label{exex}
    \lim_{v \to \infty } \frac{1}{v ^{2}}\, \ln
    \bbP \Biggl\{ \omega :\sup_{x\in \Lambda_{\ell }(0)} 
    |V ^{(\omega )}(x)|\ge v  \Biggr\} = 
    - \, \frac{1}{2C(0)} \,.
  \end{equation}
  It is valid under the assumptions of Lemma~\ref{fluctLemma} and the
  additional requirement that $C(x)\neq C(0)$ for all $x\in
  \Lambda_{\ell }(0) \setminus \{ 0\}$,
  see e.g.\  Thm.~8 in Sect.~14 of \cite{Lif95} or \cite{Ber85}. For more
  stringent, but less explicit bounds than \eqref{fluctInequ} the
  reader may consult \cite{Fer75,Lif95}. 
\end{remark}

\subsection{Initial Estimates} \label{inessec}

For the Gaussian model considered here, we are able to show spectral
localization for low energies or weak disorder. 

\begin{lemma}[Low Energies] \label{lowLemma}
  Let $\varrho, r_{0} >0$, $0<r<r_{0}$ and $N\in \bbN$
  be given and assume that the Gaussian random potential $V$ with
  covariance function $x \mapsto \sigma^2 C(x)$ has 
  property \ass{H}. Then, for every given $\sigma >0$ there is a
  length $0 <  L_{\mathsf{H}} < \infty$ such 
  that for every $L \ge L_{\mathsf{H}}$ there is $-\infty <
  \varepsilon(L) < 0$ with 
  \begin{equation}
    \bbP \big\{ \omega : V^{(\omega)} \text{ is }
    (r,E,L,x)\text{-regular} \big\} \ge 1-L^{-\varrho } 
  \end{equation}
 for all $E \in \: ]-\infty, \varepsilon(L)]$ and all $x \in
 \rz^d$. The length $L_{\mathsf{H}}$ depends on $N,r_{0}$ and on the
 H\"older exponent and the size of the neighbourhood referred to in property
 \ass{H}.  
\end{lemma}

\begin{proof}
  Let 
  \begin{equation}
    \label{esigma}
    E_{\sigma}(L):= \sigma \sqrt{200\, C(0) \ln(L) \ln\bigl(2^{2(d+1)}\,
    L^{\varrho} \bigr)}
  \end{equation}
  and set $\varepsilon(L):= -  E_{\sigma}(L) -\Delta E$ with some
  $\Delta E > 0$ fixed, but arbitrary.  
  Let $\omega\in\Omega$ such that the inequality
  \begin{equation} 
    \label{boundcond}
    \sup_{y \in \Lambda_L (x) } \bigl\{ |V^{(\omega )}(y)|\bigr\} \le
    E_{\sigma}(L)
  \end{equation}
  is satisfied. Then, Lemma \ref{boundFrame} and the Combes-Thomas
  Lemma~\ref{cotho} imply the estimate for all $E\le \varepsilon (L)$
  \begin{equation}
    \sup_{\eta>0} 
    \bignorm{R_{B_{N}^{L}(x),B_{0}^{L}(x)}(V^{(\omega)},E+\i\eta)}  
    \le
    \left( L \sqrt{v_{0}^{(\omega)}-E\, }\right)^{-r} \,
    g\left( L \sqrt{v_{0}^{(\omega)}-E\, }\right)\,.
  \end{equation}  
  Here, $v_{0}^{(\omega)} := \inf_{y\in\Lambda_L (x)} \{ V^{(\omega)}(y)\}$
  and we have defined a function $g$ on $\rz$ by 
  \begin{equation}
    g(z) := c_N\, z^{r+ (d-3)/2} 
     \left( 1 + \frac{d^{2}}{8\tilde{c}_{N}\, z} \right)
     \exp\{- \tilde{c}_N\, z\}
  \end{equation}
  with suitable constants $c_{N}, \tilde{c}_{N} < \infty$ depending
  only on 
  $N$. Observe that there exists $L_{0}\equiv L_{0}(N,r_{0},\Delta E)$
  such that $g$ satisfies the inequality 
  \begin{equation} 
    \label{finequ}
    g\bigl(\tilde{L}\sqrt{\Delta E}\bigr) \le (\Delta E)^{r/2}
  \end{equation}
  for all $\tilde{L}\ge L_{0}$ and all $0<r<r_{0}$.
  Clearly, \eqref{finequ}
  also holds with $\tilde{L}= L \sqrt{(v_{0}^{(\omega)}-E) / \Delta
    E}$, if $L\ge L_{0}$ and $E\le \varepsilon(L)$, because
  $v_{0}^{(\omega)}-E \ge \Delta E$ for all $\omega$ obeying
  \eqref{boundcond}. Therefore 
  we have $(r,E,L,x)$-regularity of $V^{(\omega)}$ 
  under this condition. But according to Lemma~\ref{fluctLemma} 
  the probability of the event \eqref{boundcond}
  is, independently of $x$, at least $1-L^{-\varrho}$ provided 
  $L\ge \ell_{\mathsf{H}}$. The proof is completed by setting
  $L_{\mathsf{H}}:=\max\{\ell_{\mathsf{H}},L_{0}\}$.
\end{proof}

\begin{lemma}[Weak Disorder] \label{smallLemma}
  Let $\varrho, r_{0} >0$, $0<r<r_{0}$ and $N\in\bbN$ be given and
  assume that the  
  Gaussian random potential $V$ with covariance function
  $x \mapsto \sigma^2 C(x)$ has property \ass{H}.
  Then, for every given energy $-\infty < E_0 < 0$ there is a length 
  $0< L_{\mathsf{H}} < \infty$ such that for every
  $ L \ge L_{\mathsf{H}} $ one can find $ \sigma(L) >0 $ with
  \begin{equation} 
    \label{smallglg}
    \bbP \big\{ \omega : V^{(\omega)} \text{ is }
    (r,E,L,x)\text{-regular} \big\} \ge 1-L^{-\varrho } 
  \end{equation}
  for all $E \in \: ]-\infty, E_0]$,
  all $x \in \rz^d$ and all $\sigma \in \: ]0,\sigma(L) ]$. 
  The length $L_{\mathsf{H}}$ depends on $N,r_{0},E_{0}$ and 
  on the H\"older exponent and the size of the neighbourhood referred
  to in property \ass{H}.
\end{lemma}

\begin{proof}
  Given $-\infty < E_{0} < 0$ and $\sigma >0$, set $\Delta E :=
  -E_{0}/2$ and $E_{\sigma}(L)$ as in \eqref{esigma}. As in the proof of
  Lemma~\ref{lowLemma} we infer the existence of a finite length 
  $L_{\mathsf{H}}:= \max\{\ell_{\mathsf{H}}, L_{0}(N, r_{0},
  -E_{0}/2)\}$, which is independent of $\sigma$, such that
  the inequality \eqref{smallglg} holds for all $E \le -E_{\sigma}(L) +
  E_{0}/2$. The proof is completed by requiring $E_{\sigma}(L) \le
  -E_{0}/2$, that is $\sigma \le \sigma(L)$ with
  \begin{equation}
    \label{sigmaell}
    \sigma(L) := -(E_{0}/2) \bigl[200\,
    C(0) \ln(L) \ln\bigl(2^{2(d+1)}\, L^{\varrho}
    \bigr)\bigr]^{-1/2}\,.
  \end{equation}
\end{proof}

\begin{remarks}
  \begin{nummer}
  \item The low-energy initial estimate, Lemma~\ref{lowLemma}, could
    have also been obtained by using the Wegner estimate
    \eqref{wegnerloc} together with the decay of $W(E)$ for
    $E\to -\infty$ instead of the Combes-Thomas estimate \eqref{coTho1}.
    But this would require property \ass{P} in addition to property
    \ass{H}. In any case, we would not know how to derive the
    weak-disorder initial estimate without the Combes-Thomas estimate.
  \item The reason why we are not able to prove a strong-disorder
    initial estimate is that controlling the frame operator 
    \eqref{frameop} with Lemma~\ref{boundFrame} requires a control of
    the excursions of the random potential $V$ to extremely negative
    energies. But these unwanted events occur with increasing
    probability if the disorder is increased, see the exact
    asymptotics \eqref{exex}. A successive application of the Wegner
    estimate \eqref{wegnerloc} instead of the Combes-Thomas estimate
    would not cure the situation either, because our Wegner constant
    \eqref{wegnerkonst} does not vanish in the strong-disorder limit.  
  \end{nummer}
\end{remarks}

\subsection{Controlling Long-Range Correlations}

In this subsection we are concerned with Assumption~\itemref{indepass} of
Theorem~\ref{multThm}, which ensures that sufficiently spatially 
separated events are sufficiently decorrelated. We will distinguish
the two different cases whether the random potential $V$ has property
\ass{M} or property \ass{R}.

In case of property \ass{M} we will apply Theorem~\ref{multThm} with
$V_{k}=V$ for all $k\in\bbN_{\,0}$ in Subsection~\ref{proofmsaThm}
below. For this purpose we will need

\begin{lemma}
  \label{Mindep}
  Let $V$ be a Gaussian random potential on
  $\rz^{d}$ with property \ass{M} and put $\vartheta_{0} := \nu/2 +
  (2/\delta) (d-1)(\nu-1) < 1$. Then, for all integers $2 \le K \le K_{0}$,
  all $\vartheta_{0} < \vartheta < 1$ and all $\rho >0$ obeying
  \begin{equation}
    \label{rhocond}
    (4/K)(d-1)(\nu-1)/(2\vartheta -\nu) < \rho \le \delta /K
  \end{equation}
  there exists a finite length $L_{\mathsf{M}} >0$ such that 
  the random potential $V$ is $(K,L,\rho,\vartheta)$-independent in
  the sense of Definition~\ref{KLindep} for all $L>L_{\mathsf{M}}$. The
  length $L_{\mathsf{M}}$ depends on $K$, $\vartheta$ and $\nu$.
\end{lemma}

\begin{remark}
  For Lemma~\ref{Mindep} to hold it is irrelevant whether $V$ is
  Gaussian or not.
\end{remark}

\begin{proof}[Proof of Lemma \ref{Mindep}]
  Let $N\in\nz$ and let $A$, $\nu$, $K_{0}$ and $\delta$ as in property
  \ass{M}. By Definition~\ref{strongmix} of the
  strong-mixing coefficient we have for all integers $2 \le K \le
  K_{0}$ and all 
  $A_{k}\in \mathcal{A}_{V}(\Lambda_{k})$, $k=1,\ldots,K$, with
  $|\Lambda_{k}| \le (L/N)^{d/\nu}$ and $\dist_{\infty}(\Lambda_{k},
  \Lambda_{k'}) \ge L/(4N)$ for $k\neq k'$ that 
  \begin{equation}
    \bbP\left(\bigcap_{k=1}^{K} A_{k}\right) \le 
    \bbP(A_{1}) \;\bbP\left(\bigcap_{k=2}^{K} A_{k}\right) + 
    \alpha_{V}\Bigl(L/(4N), (K-1)(L/N)^{d/\nu}\Bigr)\,.
  \end{equation}
  Iterating this inequality and using $\bbP(A_{k}) \le
  (L/N)^{-\rho/\nu}$, where $\rho >0$ as required in
  Definition~\ref{KLindep}, we obtain  
  \begin{multline}
    \bbP\left(\bigcap_{k=1}^{K} A_{k}\right) \le 
    (L/N)^{-K\rho/\nu} \\
    + \alpha_{V}\Bigl(L/(4N), (K-1)(L/N)^{d/\nu}\Bigr) 
    \sum_{k=0}^{K-2} (L/N)^{-k\rho/\nu}\,.
  \end{multline}
  The $k$-sum is bounded from above by $K-1$ for $L\ge N$ and
  property \ass{M} yields for all $\rho \le \delta /K$
  \begin{equation}
    \bbP\left(\bigcap_{k=1}^{K} A_{k}\right) \le 
    [1 + (K-1)A]\, (L/N)^{-K\rho/\nu}\,.
  \end{equation}
  Since $\delta > 4(d-1)(\nu-1)/(2-\nu)$, inequality \eqref{rhocond}
  is satisfied for all $\vartheta_{0} < \vartheta < 1$ 
  and \eqref{KLindepglg} of Definition~\ref{KLindep} holds
  for all $L > L_{\mathsf{M}} := N [1 + (K-1)A]^{\nu/[K\rho (1-\vartheta)]}$.
\end{proof}

In case of property \ass{R} we now construct a sequence
$\{V_{k}\}_{k\in\bbN_{\,0}}$ of Gaussian random potentials such that
$V_{k}$ satisfies Assumption~\itemref{indepass} of
Theorem~\ref{multThm} on length scale $L_{k}$ and converges suitably
to $V$.

\begin{lemma}
  \label{Rindep}
  For $N \ge 4$ integer and a monotone increasing sequence
  $\{L_{k}\}_{k\in\bbN_{\,0}}$ of length scales with $L_{0}> 8N$ and
  $\lim_{k\to \infty}L_{k} = \infty$ define the box $B_{k} :=
  \bigl(\Lambda_{L_{k}/(4N)}(0),1\bigr)$ and set $B_{\infty} :=
  (\rz^{d},0)$. Let $V$ be a Gaussian random potential on $\rz^{d}$
  with property \ass{R}. Then there exist Gaussian random potentials $V_{k}:
  \Omega\times\rz^{d} \rightarrow \rz$, $(\omega,x)\mapsto
  V_{k}^{(\omega)}(x)$, with covariance functions
  \begin{equation}\label{Ckk}
    x \mapsto C_{k}(x) := \bbE \{V_{k}(x) V_{k}(0) \} := 
    \intr[d]{y} \gamma_{k}(y) \,\gamma_k (x+y)\,,
  \end{equation}
  where $\gamma_{k} := \gamma \indfkt{B_{k}}$, such that for all
  $k\in\bbN_{\,0}$ 
  \begin{indentnummer}
    \item \label{Risoton}
      $0 < C_{0}(0) \le C_{k}(0)\le C_{k'}(0) \le C(0)$ for all $k'
      \ge k$,
    \item \label{RHoelder}
      $V_{k}$ has property \ass{H} uniformly in $k$ with the same
      H\"older exponent and the same neighbourhood as $V$,
    \item \label{RKLindep}
      $V_{k}$ is $(K,L_{k},\rho,\vartheta)$-independent in the sense
      of Definition~\ref{KLindep} for all $K \ge
      2$ integer, $\rho >0$ and $0 < \vartheta \le 1$,
    \item \label{Rapprox} 
      ~\par\vspace{-7ex}
      \begin{multline} 
        \label{Rapproxglg} 
        \bbP\left\{\omega: \sup_{x\in\Lambda_L (0)}
          |V^{(\omega)}_k (x) - V^{(\omega)}_{k'}(x)| \ge E \right\} \\
        \le 2^{2(d+1)} \exp\left\{ -\,
          \frac{E^{2} L_{k}^{2\zeta-d } }{\tilde{\gamma} \ln L}
        \right\}
      \end{multline} \par\vspace{-1ex}\noindent
      for all $k\le k'\le \infty$, all $E\ge 0$ and all $L\ge \tilde{L}$.
      Here we have set $V_{\infty}:=V$, and 
      $\tilde{\gamma}$ and $\tilde{L}$ are some strictly positive 
      constants which are independent of $k$ and $k'$.
  \end{indentnummer}
\end{lemma}

\begin{remark}
  It follows from \eqref{pointwise} below that $V_{k}(x)$ converges
  for all $x\in\rz^{d}$ to $V(x)$ in $\bbP$-mean-square sense.
  Moreover, assertion~\itemref{Rapprox} of the lemma implies that $V_{k}$
  converges also uniformly in $x$ on bounded cubes to $V$ in probability.
\end{remark}

\begin{proof}[Proof of Lemma~\ref{Rindep}]
  We denote the Fourier (-Plancherel) transform of $f\in\RL^{2}(\rz^{d})$ by 
  $\hat{f}: \rz^{d}\ni q \mapsto \hat{f}(q) :=
  (2\pi)^{-d/2}\int_{\rz^{d}}\d^{d}x\, 
  \e^{-\i q\cdot x}\, f(x)$. The matrix-valued function 
  $\rz^{d} \ni x \mapsto \mathsf{D}(x)$ with matrix elements 
  \begin{equation}\label{CkjDef}
    \mathsf{D}_{k,k'}(x) := \intr[d]{q} \e^{\i q\cdot x} \, 
    |\hat{\gamma}_k (q)|
    |\hat{\gamma}_{k'} (q)| \,,
    \qquad\quad k,k'\in \bbN_{\,0}\cup\{\infty\}\,,
  \end{equation}
  is continuous in $x$ due to 
  $|\hat{\gamma}_{k}| |\hat{\gamma}_{k'}| \in \RL^{1}(\rz^{d})$, which
  follows from $\gamma_{k} \in \RL^{2}(\rz^{d})$ and thus
  $\hat{\gamma}_k \in \RL^{2}(\rz^{d})$. Each $\mathsf{D}_{k,k'}$ is
  a real-valued and even function in $x$, because $|\hat{\gamma}_{k}(q)|^{2} = 
  \hat{\gamma}_{k}(q) \hat{\gamma}_{k}(-q)$. Moreover, since each
  $\mathsf{D}_{k,k'}$ has a non-negative Fourier transform
  which factorizes in $k$ and $k'$, the Bochner-Khintchine theorem
  implies that $x \mapsto \mathsf{D}(x)$ is a matrix-valued covariance
  function. Therefore
  -- see e.g.\ Thm.~4.2 in \cite{Lif95} or Appendix~A.4 to Part~I of
  \cite{GlJa87} --,   
  there exists a sequence $\{V_{k}\}_{k\in\bbN_{\,0}\cup\{\infty\}}$ of
  \emph{jointly} Gaussian homogeneous random fields on some complete
  probability space $(\Omega',\mathcal{A}',\bbP\mkern2mu')$ with
  zero mean and covariance $\bbE\{ V_{k}(x) V_{k'}(0)\} =
  \mathsf{D}_{k,k'}(x)$. 
  Without loss of generality we assume that the original probability
  space $(\Omega,\mathcal{A},\bbP)$ underlying $V$ coincides with
  $(\Omega',\mathcal{A}',\bbP\mkern2mu')$. The
  same arguments as in the proof of Lemma~\ref{gauss1.2.4} show that
  each $V_{k}$ has a separable and jointly measurable version.
  We only consider these versions. Since the non-negative function
  $\gamma$ is determined by \eqref{Frep} only up to Euclidean translations, we
  can assume without loss of generality that $\gamma(0) > 0$. Thus
  we have $0 < C_{k}(0) \le C(0)$ for the covariance function 
  $C_{k}(x):= \mathsf{D}_{k,k}(x)$ of $V_{k}$ for all
  $k\in\bbN_{\,0}\cup \{\infty\}$. Moreover, the sequence
  $\{C_{k}(x)\}_{k\in\bbN_{\,0}}$ is increasing  
  for all $x\in\rz^{d}$. This proves that $\{V_{k}\}_{k\in\bbN_{\,0}}$ is a
  sequence of Gaussian random potentials on $\rz^{d}$ in the sense of
  Definition~\ref{Gaussrp} whose covariance functions satisfy 
  assertion~\itemref{Risoton} of the lemma. In particular, the
  relation $C_{\infty}=C $ allows one to identify $V_{\infty}$ and $V$.

  As to Assertion~\itemref{RHoelder}, we remark that for all $k\in\bbN_{\,0}$
  an indicator function $\indfkt{B_{k}}\in
  \mathcal{C}_{0}^{\infty}(\rz^{d})$ of the box $B_{k}$ is uniformly H\"older
  continuous with exponent one, 
  \begin{equation}
    \bigl| \indfkt{B_k} (x+y) -
      \indfkt{B_k} (x) \bigr| 
    \le 
    d \varkappa_{1}\, |y|_{\infty} \qquad\quad
    \text{for all } x,y\in\rz^{d}\,,
  \end{equation}
  as a consequence of the mean-value theorem and the boundedness 
  \eqref{kappa1}. Note that the constant $\varkappa_{1}$ does not
  depend on $k$. Therefore we get, uniformly in $k$,
  \begin{equation}
    \label{uchi}
    | \gamma_k(x+y) - \gamma_k(x) | 
    \le 
    \left( \gamma_{0}\, d\varkappa_{1} \sup_{y\in
        U}\{|y|_{\infty}^{1-\alpha} \} + a \right) 
    |y|_{\infty}^{\alpha}
  \end{equation}
  for all $x\in\rz^{d}$ and all $y\in U$, the 
  neighbourhood of the origin referred to in property \ass{R} for $V$. 
  Hence, $\gamma_{k}$ is uniformly H\"older continuous for
  all $k\in\bbN_{\,0}$ with the same exponent $\alpha$ and the same
  neighbourhood of the origin as given for $\gamma$.

  Assertion \itemref{RKLindep} follows from $\gamma_{k}(x) = 0$
  for all $|x|_{\infty} \ge L_{k}/(8N)$. Hence, $C_{k}(x)=0$ 
  for all $|x|_{\infty} \ge L_{k}/(4N)$ by \eqref{Ckk}, and the
  Gaussian nature of $V_{k}$ implies the stochastic independence of events
  which are at least a distance $L_{k}/(4N)$ apart.

  Assertion \itemref{Rapprox} will be deduced from an
  application of Lemma~\ref{fluctLemma} to the left-hand side of
  \eqref{Rapproxglg}. To this end observe that by construction the
  difference $V_{k} -V_{k'}$ is also 
  a Gaussian random potential on $\rz^{d}$ for all 
  $k,k' \in \bbN_{\,0} \cup\{\infty\}$. Its covariance function  
  \begin{equation}
    \label{diffCov}
    \bbE\left\{
      \big( V_k(x) - V_{k'}(x) \big) \big( V_k(0) - V_{k'}(0) \big)
    \right\} =
    C_{k}(x)+C_{k'}(x)-2 \mathsf{D}_{k,k'}(x) 
  \end{equation}
  has property \ass{H} with the same H\"older exponent $\alpha$ and
  the same neighbourhood  as given for the covariance function of $V$. This
  follows from assertion~\itemref{RHoelder} and $\mathsf{D}_{k,k'}(0) -
  \mathsf{D}_{k,k'}(x) \ge 0$ for all $x\in\rz^{d}$. Moreover, since
  the Fourier transformation is an isometry on $\RL^{2}(\rz^{d})$, we
  conclude for $k<k'$ the inequality
  \begin{equation} \label{pointwise}
    \bbE \bigl\{\bigl( V_{k}(0) - V_{k'}(0)\bigr)^{2}\bigr\}
    \le  \norm{\widehat{(\gamma_k - \gamma_{k'})}}^{2}  \!
    = \bignorm{\gamma_k- \gamma_{k'}}^{2} 
    \le \frac{\tilde{\gamma}}{200}\, L_{k}^{d-2\zeta }
  \end{equation}
  with some $k,k'$-independent constant $0< \tilde{\gamma}< \infty$
  and $\zeta$ taken from property \ass{R}. 
  Hence, \eqref{Rapproxglg} follows from Lemma~\ref{fluctLemma}
  for all $L \ge \tilde{L}=\ell_{\mathsf{H}}$ with $\tilde{L}$ being also
  independent of $k$ and $k'$ due to the uniform H\"older property
  \itemref{RHoelder} of  $V_{k} - V_{k'}$. 
\end{proof}

\subsection{Proof of Theorem \ref{msaThm}} \label{proofmsaThm}

We are now in a position to prove Theorem~\ref{msaThm}. This proof
also completes the proof of the main Theorem~\ref{maintheo}.

\begin{proof}[Proof of Theorem \ref{msaThm}]
  We show that Theorem~\ref{msaThm} follows from an application of 
  Theorem~\ref{multThm}. To do so we have to check the assumptions of
  this theorem. 

  Let $N=4$ and $S=2$. Choose the exponents $\nu$ and $r$ such that
  \begin{gather}
    1 < \nu < 1 + \frac{1}{8d}\,, \label{nubed}\\
    4d\nu < r < 4d +\frac{1}{2} =:r_{0} \label{rbed}
  \end{gather}
  in accordance with \eqref{msaNu} and \eqref{rNuD}. The precise value
  of $\nu$ will be fixed later on. In case that $V$
  has properties \ass{P}, \ass{H} and \ass{M} -- let us call this
  simply case \ass{PHM} -- we set $V_{k} = V$ for all
  $k\in\bbN_{\,0}$. In case that $V$ has property \ass{R} we use the
  sequence of Gaussian random potentials constructed in
  Lemma~\ref{Rindep} for the sequence \eqref{lenseq} of length scales
  $L_{k}$.  For the low-energy assertion~\itemref{lowpart} of
  Theorem~\ref{msaThm} we consider $\sigma >0$ arbitrary, but fixed,
  and set $I_{k} = ]-\infty , \varepsilon (L_{k})]$ in
  Theorem~\ref{multThm}, $\varepsilon (L)$ being the energy defined in
  Lemma~\ref{lowLemma} and given explicitly below \eqref{esigma}. For the
  weak-disorder assertion~\itemref{weakpart} 
  of Theorem~\ref{msaThm} we set $I_{k} = ]-\infty , E_{0}]$ for all 
  $k\in\bbN_{\,0}$, but allow only for disorder strengths $\sigma \in
  ]0, \sigma(L_{k})]$ on the length scale $L_{k}$ with $\sigma(L)$ given
  by \eqref{sigmaell} in the proof of Lemma~\ref{smallLemma}. For the time
  being we suppose that the initial length 
  satisfies $L_{0} > \max\{2^{2/(\nu -1)}, L_{\mathsf{H}},
  L_{\mathsf{M}}\}$. The length $L_{\mathsf{H}}$ is to
  be taken from Lemma~\ref{lowLemma} for the low-energy assertion,
  respectively from Lemma~\ref{smallLemma} for the weak-disorder
  assertion. The length $L_{\mathsf{M}}$ was defined in
  Lemma~\ref{Mindep}.

  Hence, considering case \ass{PHM}, we conclude that Lemma~\ref{lowLemma},
  respectively Lemma~\ref{smallLemma}, 
  provides the Initial-Estimate Assumption~\ref{initass} for
  assertion~\itemref{lowpart}, respectively assertion~\itemref{weakpart}, of 
  Theorem~\ref{msaThm} for any $0<r<r_{0}$ and $\rho >0$. The same is
  true for case \ass{R}, because
  Lemma~\ref{Risoton} and \itemref{RHoelder} ensure that 
  Lemmas \ref{lowLemma} and \ref{smallLemma} remain true with the same
  constants $L_{\mathsf{H}}$, $\varepsilon(L)$ and $\sigma(L)$ of case
  \ass{PHM}, if $V$ is replaced by $V_{k}$.

  Next we come to Assumptions~\ref{vorindepass} and
  \itemref{indepass}. First, we consider the case \ass{PHM} and fix 
  $\nu$ as required by property \ass{M}. Obviously, both assumptions
  then follow from Lemma~\ref{Mindep} with $K=N-S=2$. Moreover, observing
  \eqref{nubed} and choosing $\vartheta > \max\{3/4,\vartheta_{0}\}$,
  the left inequality in  
  \eqref{rhocond} guarantees that Assumptions~\ref{vorindepass} and
  \itemref{indepass} hold for some 
  \begin{equation}
    \label{rhomaxbed}
    \rho < 1/2\,.
  \end{equation}
  Concerning case \ass{R}, Assumption~\ref{indepass} holds for all $0
  < \vartheta \le 1$ and all $\rho >0$ by
  Lemma~\ref{RKLindep}. Thus we may also pick $\nu$, $\rho$ and
  $\vartheta$ as in case \ass{PHM}, thereby satisfying
  Assumption~\ref{vorindepass} and \eqref{rhomaxbed}.

  Taking into account \eqref{nubed} and \eqref{rbed}, it follows that
  a sufficient condition for Assumption~\ref{firstass} to hold is
  \begin{equation}
    \label{wmaxbed}
    w < d - 1/4\,.
  \end{equation}
  We will show below that in case \ass{PHM}
  Assumption~\ref{nonruappass} is satisfied 
  for all $\rho,w >0$ which obey 
  \begin{equation}
    \label{wminbed}
    w > \rho + d + 1 - \frac{2}{\nu}\,.
  \end{equation}
  We will also show below that Assumption~\ref{nonruappass} is
  satisfied in case \ass{R} if in addition to \eqref{wminbed} the
  inequality  
  \begin{equation}
    \label{zetaminbed}
    \zeta > \frac{d}{2} + r + 2w + \frac{2}{\nu} -1
  \end{equation}
  is assumed for the exponent $\zeta$.
  Since Assumptions~\ref{vorindepass} and \itemref{indepass} can be
  fulfilled for some $\rho < 1/2$  and
  since $\nu$ obeys \eqref{nubed}, it follows that \eqref{wmaxbed} and
  \eqref{wminbed} are compatible. Moreover, since property \ass{R}
  requires $\zeta > 13d/2 + 1$, the condition \eqref{zetaminbed} does
  not impose a further restriction beyond \eqref{nubed}, \eqref{rbed}
  and \eqref{wmaxbed}. Up to now we have shown that Assumptions
  \ref{firstass}, \itemref{vorindepass}, \itemref{indepass} and
  \itemref{initass} can be satisfied under the conditions
  \eqref{wminbed} and \eqref{zetaminbed}. Thus, in order to
  complete the proof it remains to show that \eqref{wminbed} and
  \eqref{zetaminbed} are sufficient conditions for
  Assumption~\ref{nonruappass} to hold.

  We consider both cases \ass{PHM} and \ass{R} simultaneously. 
  Since $V_{k} = V$ for all $k$ in case \ass{PHM}, there is less to
  prove in case \ass{PHM} than in case \ass{R}.
  Let $E\in I_{0}$, $x\in\rz^{d}$ and $U=V$ or $U=V_{k+1}$. Then, the 
  inequality 
  \begin{align} \label{pmonster}
    \bbP & \big\{\omega :\,   V^{(\omega )}_{k} \text{ is }
    (w,E,L_{k},x)\text{-non-resonant and} \nonumber\\ 
    & \qquad\text{an }(r,E,L_{k},x)\text{-approximation of } U^{(\omega
      )}\; \big\} \nonumber\displaybreak[0]\\
    & \ge   \bbP\Bigg\{ \omega : \varPhi_{B_{n}^{L_{k}}(x)}(E -
    V_{k}^{(\omega)}) < g_{n} \text{~~for all~} 1\le n \le 4
    \displaybreak[0]\nonumber\\ 
    & \qquad\qquad \text{and\;} \dist\Bigl(E, \mathrm{spec}\bigl(
    H_{\Lambda_{nL_{k}/4}(x)}(V_{k}^{(\omega)})\bigr)\Bigr) \ge
    g_{n} L_{k}^{-w} \text{~~for all~} 1\le n \le 4
    \displaybreak[0]\nonumber\\  
    & \qquad\qquad \text{and\;} \dist\Bigl(E, \mathrm{spec}\bigl(
    H_{\Lambda_{L_{k}}(x)}(U^{(\omega)})\bigr)\Bigr) \ge
    L_{k}^{-w'} \nonumber\\ 
    & \qquad\qquad \text{and\;} \bignorm[\infty;
    \Lambda_{L_{k}}(x)]{V_{k}^{(\omega)} - U^{(\omega)}} 
    \Biggl( L_{k}^{w} + \frac{16\sqrt{2}\,\varkappa_{1}}{L_{k}
      \sqrt{g_{4}}}\, L_{k}^{w/2}\Biggr) \le \frac{L_{k}^{-w'-r}}{2}
     \Bigg\}
  \end{align}
  is valid for all $w'>0$ and all $g_{n}>0$, $n=1,\ldots,4$. It
  follows from Definition~\ref{nonrdef}, Definition~\ref{appdef} and the
  inequalities 
  \begin{gather*}
    \begin{align}
      \bignorm{R_{B_{n}^{L_{k}}(x), B_{0}^{L_{k}}(x)} (V_{k},
        E+\i\eta)} & \le \varPhi_{B_{n}^{L_{k}}(x)}(E-V_{k})\,
      \bignorm{\bigl( H_{\Lambda_{nL_{k}/4}(x)}(V_{k}) - E\bigr)^{-1}}
      , \nonumber \\ \\[.2ex]
      \bignorm{W_{B_{n'}^{L_{k}}(x), B_{4}^{\ell_{k}}(y)} (V_{k},
        E+\i\eta)} & \le \varPhi_{B_{n'}^{L_{k}}(x)}(E-V_{k})\,
      \bignorm{\bigl( H_{\Lambda_{n'L_{k}/4}(x)}(V_{k}) -
        E\bigr)^{-1}}, \nonumber \\
    \end{align} \\[.2ex]
    \begin{align}  
      \bignorm{D_{x}^{L_{k}}(U,V_{k},E+\i\eta)}  \le &  \bignorm[\infty;
      \Lambda_{L_{k}}(x)]{V_{k} - U} \, \bignorm{\bigl(
        H_{\Lambda_{L_{k}}(x)}(U) - E\bigr)^{-1}} \nonumber\\
      & \times\Bigg\{ \varPhi_{B_{4}^{L_{k}}(x)}(E-V_{k})\, 
      \bignorm{\bigl( H_{\Lambda_{L_{k}}(x)}(V_{k}) - E\bigr)^{-1}}
      \nonumber\\
      & \hspace*{1.3cm} + \frac{16\sqrt{2}\,\varkappa_{1}}{L_{k}}\;
      \bignorm{\bigl( H_{\Lambda_{L_{k}}(x)}(V_{k}) -
        E\bigr)^{-1}}^{1/2}  \Bigg\}\,, \nonumber\\
    \end{align}
  \end{gather*}
  where $1\le n < n'\le 4$, $y\in T_{n}^{L_{k}}(x)$, $\ell_{k}:=
  (L_{k}/4)^{1/\nu}$. To derive the latter three inequalities  
  Lemma~\ref{boundFrame} and \eqref{RFrame} have been used. Note that
  in case \ass{PHM} the last two lines of \eqref{pmonster} have simply
  to be omitted. Since $\bbP(A
  \cap B) \ge \bbP(A) - \bbP(\Omega\!\setminus\! B)$ for $A,B\in
  \mathcal{A}$, we bound the right-hand side of \eqref{pmonster} from
  below by 
  \begin{align} \label{vorvor}
    \bbP\Bigl\{ & \omega :\,  \varPhi_{B_{n}^{L_{k}}(x)}(E -
    V_{k}^{(\omega)}) < g_{n} \text{~~for all~} 1\le n \le 4
    \Bigr\} \nonumber\\ 
    & - \;\sum_{n=1}^{4} \bbP\Bigl\{ \omega :\,  \dist\Bigl(E,
    \mathrm{spec}\bigl( H_{\Lambda_{nL_{k}/4}(x)}
    (V_{k}^{(\omega)})\bigr)\Bigr)  
    < g_{n} L_{k}^{-w}\Bigr\} \nonumber\\
    & - \;\bbP\Bigl\{ \omega :\,  \dist\Bigl(E,
    \mathrm{spec}\bigl( H_{\Lambda_{L_{k}}(x)}
    (U^{(\omega)})\bigr)\Bigr) < L_{k}^{-w'}\Bigr\} \nonumber\\
    & - \;\bbP\Bigg\{ \omega :  \bignorm[\infty;
    \Lambda_{L_{k}}(x)]{V_{k}^{(\omega)} - U^{(\omega)}} >
    \frac{L_{k}^{-(w' +w +r)}}{2} 
    \left(\! 1 \!+ \!
      \frac{16\sqrt{2}\,\varkappa_{1}}{L_{k}^{1+w/2}\sqrt{g_{4}}}
      \right)^{-1} \Bigg\}\,.
  \end{align}
  Choosing $g_{n} = \varPhi_{B_{n}^{L_{k}}(x)}
  \bigl(E-v_{0}(L_{k})\bigr)$ with 
  \begin{equation}
    v_{0}(L):= - \sigma \sqrt{
      200\,C(0) \ln(L) \ln\bigl(2^{2(d+2)}L^{\varrho}\bigr)\,}\,,
  \end{equation}
  the first probability in \eqref{vorvor} is seen to be bounded from below by
  $\bbP\bigl\{\omega :\,\bignorm[\infty; \Lambda_{L_{k}}
  (x)]{V_{k}^{(\omega)}} < -v_{0}(L_{k}) \bigr\} \ge 1- L_{k}^{-\rho}/4$ by
  virtue of Lemma~\ref{fluctLemma}. The probabilities in the second
  and third line of \eqref{vorvor} are estimated from above with the
  Wegner bound \eqref{wegnerloc} and the last probability is estimated
  from above with Lemma~\ref{Rapprox}. Thus, \eqref{vorvor} is bounded
  from below by 
  \begin{align}
    \label{vorletzte}
    1 & - \;\frac{L_{k}^{-\rho}}{4}\;  -
    2W_{0}\sum_{n=1}^{4}\left(\frac{n}{4}\right)^{d} g_{n}\,
    L_{k}^{d-w} - 2W_{0}\, L_{k}^{d-w'} \nonumber\\
    & - 2^{2(d+1)} \exp\left\{ - \,\frac{L_{k}^{2( \zeta -d/2 - w' -w
          -r)}}{4 \tilde{\gamma}\, \ln L_{k}}  
    \left( 1 + 
      \frac{16\sqrt{2}\,\varkappa_{1}}{L_{k}^{1+w/2}\sqrt{g_{4}}}
      \right)^{-2}\right\}\,.
  \end{align}
  The constant $W_{0} >0$ is an upper bound on the arising
  Wegner constants which is uniform in $k$ due to
  Lemma~\ref{Risoton}, \eqref{wegnerkonst} and because the subset 
  $\{ E + g_{n} L_{k}^{-w} + L_{k}^{-w'} : E\in I_{0}, 1\le n \le 4, k
  \in \bbN_{\,0}\}$ of the real line has a finite supremum. The
  latter is true because of \eqref{nubed} and the estimate 
  \begin{equation} \label{gup}
    g_{n}= \varPhi_{B_{n}^{L_{k}}(x)} \bigl(E-v_{0}(L_{k})\bigr) \le 
    \frac{16^{2}}{(L_{k}/4)^{2/\nu}} \left( \frac{\varkappa_{2}}{2} +
      \sqrt{\frac{\varkappa_{4}}{2} +
        2\,\frac{L_{k}^{2}\varkappa_{1}^{2}}{16^{2}}\;
        |v_{0}(L_{k})|}\;\right), 
  \end{equation}
  where we have used the definition \eqref{gfunctional} of the
  functional $\varPhi$. Again by \eqref{gfunctional}, it follows that 
  \begin{equation}
    \label{glow}
    \sqrt{g_{4}} \ge 16 \sqrt{\varkappa_{2}/2}\; L_{k}^{-1}\,.
  \end{equation}
  Combining the inequalities \eqref{pmonster}, \eqref{vorvor} and 
  \eqref{vorletzte}, choosing $w' = w -1 + 2/\nu$ and observing
  \eqref{gup}, \eqref{glow}, \eqref{wminbed} and \eqref{zetaminbed},  we
  infer the   
  existence of a finite length $L_{\mathsf{R}} >0$ such that 
  \begin{align} \label{ellerr}
    \bbP & \big\{\omega :\,   V^{(\omega )}_{k} \text{ is }
    (w,E,L_{k},x)\text{-non-resonant and} \nonumber\\ 
    & \qquad\text{an }(r,E,L_{k},x)\text{-approximation of } U^{(\omega
      )}\; \big\} \ge  1 - L_{k}^{-\rho}  
  \end{align}
  for all $L_{k}> L_{\mathsf{R}}$.  This establishes Assumption
  \ref{nonruappass} and thus completes the proof of the theorem.
\end{proof}


\appendix
\section*{An Explicit Combes-Thomas Estimate}
\addcontentsline{toc}{section}{Appendix: A Combes-Thomas Estimate}

The initial estimates of the multi-scale analysis are obtained in
Subsection \ref{inessec} with the
help of a norm estimate for the ``localized'' finite-volume resolvent at
energies below the bottom of the spectrum. In the literature such type
of estimates go with the names of J.-M.~Combes and L.~E.~Thomas
\cite{CoTh73}, see also \cite{BaCo97b} and \cite{Sto00}.

\begin{lemma} \label{cotho}
  Let $v\in\RL^{\infty}(\Lambda)$ be a bounded potential on the
  bounded open cube $\Lambda \subset\rz^{d}$. 
  Let $\Lambda_{1},\Lambda_{2}\subset\Lambda$ be disjoint Borel subsets
  of $\Lambda$ such that $\delta:=\dist(\Lambda_1,\Lambda_2) > 0$ and
  let $f_{j}: 
  \Lambda \rightarrow [0,1]$ be a measurable function with support in
  $\Lambda_{j}$ for $j=1,2$.
  Then one has for all $E<v_{0} := \essinf_{x\in\Lambda}\{v(x)\}$ the
  inequality 
  \begin{multline}\label{coTho1}
    \bignorm{ f_1 \bigl(H_{\Lambda}(v)-E \bigr)^{-1}f_2 }
    \le 
    \frac{\sqrt{|\Lambda_1||\Lambda_2|\,}}{2^{(d+1)/4}(\pi\delta)^{(d-1)/2}}  
    \; (v_{0} -E)^{(d-3)/4}\\
    \times\, \left(1+ \frac{d^{2}}{8\delta\sqrt{2(v_{0}-E)}} \right)
    \exp\left\{- \delta\sqrt{2(v_{0}-E)}\right\}\,,
  \end{multline}
  where $H_{\Lambda}(v)$ is defined on $\RL^{2}(\Lambda)$ as in
  \eqref{diriop}. 
\end{lemma}

\begin{proof}
  Let $\varphi\in\RL^{2}(\Lambda)$. From the Feynman-Kac-It\^{o}
  representation of the Schr\"odinger semigroup
  $\e^{-tH_{\Lambda}(v)}$, $t > 0$, see e.g.\ \cite{BrHuLe00}, and
  the explicit form of the heat kernel one obtains the inequality 
  \begin{multline} 
    \label{appstart}
    \bignorm{ f_{1}\,\e^{-tH_{\Lambda}(v)}f_2\,\varphi }^{2} \\ \le 
    \e^{-2tv_{0}} \intr[d]{x} \bigl(f_{1}(x)\bigr)^{2} \left(
      \intr[d]{y} \frac{e^{-(x-y)^{2}/(2t)}}{(2\pi t)^{d/2}}\;
      f_{2}(y) \, |\varphi(y)|\right)^{2} \,.  
  \end{multline}
  The fact that $f_{1}$ and $f_{2}$ are supported in spatially
  separated regions allows one to bound the right-hand side of
  \eqref{appstart} by
  \begin{equation}
    \e^{-2tv_{0}}\;\frac{e^{-\delta^{2}/t}}{(2\pi t)^{d}}\; 
    \norm{f_{1}}^{2} \langle f_{2}, |\varphi|\rangle^{2} \,.
  \end{equation}
  Estimating the scalar product by the Schwarz inequality and observing
  $\norm{f_{j}}^{2} \le |\Lambda_{j}|$, we conclude for the operator norm
  \begin{equation} 
    \bignorm{ f_1 \,\e^{-tH_{\Lambda}(v)}f_2}
    \le 
    \sqrt{|\Lambda_1||\Lambda_2|}
    \;\frac{\e^{-t v_{0}}\,\e^{-\delta^{2}/(2t)}}{(2\pi t)^{d/2}}
    \,.
  \end{equation} 
  Therefore, by Laplace transforming
  $\e^{-tH_{\Lambda}(v)}$ with respect to $t$, we get  the inequality
  \begin{multline}
    \bignorm{ f_1 \bigl(H_{\Lambda}(v)-E \bigr)^{-1}f_2}
    \le 
    \frac{2^{1/2 +d/4}\sqrt{|\Lambda_{1}||\Lambda_{_{2}|}}}{(2\pi)^{d/2}}
    \left(\frac{\delta}{\sqrt{v_{0}
        -E}}\right)^{1-d/2} \\
  \times K_{1-d/2}\left( \delta
    \sqrt{2(v_{0} -E)}\right)\,,
  \end{multline}
  where $E< v_{0}$, $K_{\nu}$ denotes the modified Bessel function of
  the second kind with index $\nu$, and formula 
  3.471.9 in \cite{Kratz} has been used. Upon inserting the series
  expansion 8.451.6 in \cite{Kratz} for $K_{\nu}$, truncating it after
  the zeroth term and estimating the remainder, we arrive at
  \eqref{coTho1}.  
\end{proof}


\acknowledgments
\addcontentsline{toc}{section}{Acknowledgments}
We would like to express our gratitude to Jean-Marie Barbaroux,
Kurt Broderix$\,$($\dagger$), Jean-Michel Combes, Peter Hislop, Dirk Hundertmark,
Thomas Hupfer, Werner Kirsch, 
Peter Stollmann, G\"unter Stolz and Simone Warzel,
with whom we enjoyed fruitful and stimulating discussions.
H.L.\ is grateful to Ilya Ya.\ Goldsheid for encouraging him to
continue research on this project.
W.F.\ thanks Jean-Michel Combes for hospitality and support at the
Centre de Physique Th\'eorique, CNRS, Marseille and at the
Universit\'e de Toulon et du Var.
P.M.\ gratefully acknowledges Annette Zippelius for travel support.
This work was supported by the Deutsche Forschungsgemeinschaft under
grant nos.\  Le~330/10-1 and Le~330/12-1. The latter is a project
within the Schwerpunktprogramm ``Interagierende stochastische Systeme
von hoher Komplexit\"at''.

\citationindex

\end{document}